\documentclass[preprint]{aastex63}
\usepackage{amsmath}
\usepackage{apjfonts}
\begin{document}

\title{The Strikingly Metal-rich Halo of the Sombrero Galaxy\footnote{Based on
    observations made with the NASA/ESA Hubble Space Telescope, obtained at the
    Space Telescope Science Institute, which is operated by the Association of
    Universities for Research in Astronomy, Inc., under NASA contract NAS
    5-26555.  These observations are associated with program GO-14175.}} 

\correspondingauthor{Roger E. Cohen}
\email{rcohen@stsci.edu}

\author{Roger E. Cohen}
\affiliation{Space Telescope Science Telescope Institute, 3700 San Martin
  Drive, Baltimore, MD 21218, USA}

\author{Paul Goudfrooij}
\affiliation{Space Telescope Science Telescope Institute, 3700 San Martin
  Drive, Baltimore, MD 21218, USA}

\author{Matteo Correnti}
\affiliation{Space Telescope Science Telescope Institute, 3700 San Martin
  Drive, Baltimore, MD 21218, USA}

\author{Oleg Y. Gnedin}
\affiliation{Department of Astronomy, University of Michigan, Ann Arbor, MI
  48109, USA}

\author{William E. Harris}
\affiliation{Department of Physics \& Astronomy, McMaster University, Hamilton,
  ON L8S\,4M1, Canada}

\author{Rupali Chandar}
\affiliation{Department of Physics \& Astronomy, University of Toledo, Toledo, OH 43606, USA}

\author{Thomas H. Puzia}
\affiliation{Institute of Astrophysics, Pontificia Universidad Cat\'olica de
  Chile, Avenida Vicu\~na Mackenna 4860, 7820436 Macul, Santiago, Chile}

\author{Rub\'en S\'anchez-Janssen}
\affiliation{STFC UK Astronomy Technology Centre, The Royal Observatory
  Edinburgh, Blackford Hill, Edinburgh, EH9\,3HJ, UK}

\begin{abstract}

The nature of the Sombrero galaxy (M\,104 = NGC 4594) has remained elusive despite
many observational studies at a variety of wavelengths.  Here we present
\textit{Hubble Space Telescope} imaging of two fields at $\sim$16 and 33 kpc
along the minor axis to examine stellar metallicity gradients in the extended
spheroid.  We use this imaging, extending more than 2 mag below the tip of the
red giant branch (TRGB), in combination with artificial star tests to forward
model observed color-magnitude diagrams (CMDs), measuring metallicity
distribution functions (MDFs) at different radii along the minor axis.
An important and unexpected result is that the halo of the Sombrero is
strikingly metal-rich: even the outer field, located at $\sim$\,17 effective
radii of the bulge, has a median metallicity ${\rm [Z/H]} \sim -0.15$ and the
fraction of stars with ${\rm [Z/H]} < -1.0$ is negligible.  This is
unprecedented among massive galaxy halos studied to date, even among giant
ellipticals. 
We find significant radial metallicity gradients, characterized by an increase
in the fraction of metal-poor stars with radius and a gradient in median
metallicity of $\sim$0.01 dex/kpc.  The density profile is well-fit by power
laws with slopes that exhibit a dependence on metallicity, with flatter slopes
for more metal-poor stars.  
We discuss our results in the context of recent stellar MDF studies of other
nearby galaxies and potential formation scenarios for the Sombrero galaxy.  

\end{abstract}

%\keywords{editorials, notices --- 
%miscellaneous --- catalogs --- surveys}

\section{Introduction} \label{sec:intro}

Stellar population gradients in the external regions of galaxies serve as
powerful diagnostic tools to constrain their formation histories.  Massive
early-type galaxies were found to be consistent with the global relationship
between halo metallicity and luminosity found for disk galaxies
\citep{mouhcine2}, indicating that such a relationship persists regardless of
galaxy type.  Recent simulations predict a halo mass-metallicity relation
\citep{auriga,illustris}, confirmed by observations of Milky Way-mass disk
galaxies despite a rich diversity in the resolved stellar population properties
of their halos \citep{harmsen}.  For these late-type galaxies, the simulations
reveal that such a correlation is a consequence of the mass-metallicity
relation of the disrupted dwarf satellites contributing to their halos,
consistent with the detection of substructure.  Furthermore, density and
metallicity gradients (and their scatter) encode information about the mass and
time of accretion of the dominant progenitor \citep{illustris}.  In particular,
galaxies with fewer significant progenitors have more massive halos and steeper
negative halo metallicity gradients and density profiles \citep{auriga}.   

Such a plethora of recent insights into the assembly histories of late-type
galaxies was facilitated by \emph{Hubble Space Telescope} (HST) imaging
campaigns of their halos \citep[e.g.][]{mouhcine3,ghosts}, resolving individual
stars on the red giant branch (RGB) to measure MDFs photometrically.  However,
for massive early-type galaxies, imaging suitable for resolved stellar
population studies of their halos is still scarce, and confined to NGC 5128
(Cen A) \citep[][and references therein]{rejkuba_5128} plus four additional E
and S0 galaxies  
\citep{harris_3377,harris_3379,sombrero_wfpc2,peacock_lent_mdf,jang_3379}.
Information on density and metallicity gradients in massive early-type galaxies
is crucial for comparisons to models, which predict a higher accreted mass
fraction than late type galaxies at fixed mass as well as a fraction of
accreted material which increases with total galaxy mass
\citep{cooper17,illustris}.  In particular, 
simulations predict that the duration of accretion is a
function of galaxy mass such that massive early-type galaxies have accreted
most of their material by $z$$\sim$2 \citep{oser10}.   

Observations of massive early-type galaxies thus far appear to be in accord
with the two-phase formation scenario and its dependence on mass. This is
evidenced by uniformly old ages for more massive early-type galaxies
\citep{int_spec_mdfs,atlas3d2,atlas3d,muse_3115}, along with shallower
metallicity gradients compared to both lower mass early-type galaxies
\citep{sluggs_met_gradient,massive} and more massive late-type galaxies
\citep{goddard_manga,manga}.  However, nearly all of these observations,
employing integrated light,  
are restricted to the inner few effective radii ($R_{\rm eff}$) of their target
galaxies (see, e.g.,~Table 5 of \citealt{goddard_manga}), whereas accreted
material likely dominates only beyond $\sim$20 kpc \citep{cooper}.  Indeed, the
HST imaging campaign of the halo of NGC 5128 at increasingly large distances
\citep[][and references therein]{rejkuba_5128} demonstrated that the halos of
even the most massive early-type galaxies can have a metal-poor contribution,
but such a contribution only becomes detectable at very large radii.

The Sombrero galaxy (= M104 = NGC 4594) is an enigmatic stellar system that has
often been considered to exhibit an archetypal classical merger-built bulge
\citep[][and references therein]{kk04}.  It is both nearby ($d$ = 9.55 Mpc;
\citealt{sombrero_trgb}) and massive, with a total stellar mass of
$\sim$2$\times$10$^{11}$M$_{\sun}$ \citep[e.g.][]{tempel,jardel}. It is viewed nearly
edge-on (inclination$\sim$84$^{\circ}$;  \citealt{emsellem}), displaying
spectacular dust lanes in the plane of a disk hosting very low-level
(0.1$M_{\odot}$/yr) star formation \citep{lwh07}.   Formally classified as type
Sa \citep{rc3}, the nature of the Sombrero galaxy is currently controversial,
due largely to the extended spheroid revealed by 3.6$\mu$ imaging from the
\textit{Spitzer Space Telescope}, in addition to multiple inner rings or disks.
In particular, two-dimensional multi-component fits to the \textit{Spitzer}
imaging by \citet[][hereafter GSJ12]{gadotti} were significantly improved when an exponential
halo was included as a third component in addition to a bulge and disk.  The
inclusion of the halo as a third component drastically affects the
structural parameters of the system, lowering the effective radius of the bulge
by a factor of seven to 0.46 kpc (along the major axis), halving the bulge
S{\'e}rsic index from $n \sim$ 4 to 2, and decreasing the bulge to total ratio
$B/T$ from 0.77 to 0.13.   

These dramatic changes are particularly intriguing when placed against the
backdrop of known scaling relations for different galaxy types.  When the
spheroid is included in the fit, the location of the bulge in the mass-size
relation becomes inconsistent with elliptical galaxies, lying much closer to the
loci of bulges and pseudo-bulges, although in both cases it is somewhat small
for its mass.  On the other hand, the smaller bulge from the three-component
fit, taken at face value, would imply a black hole mass discrepant by about an
order of magnitude from the relation between bulge mass and black hole mass, but
including the spheroid in the mass estimate alleviates this discrepancy
(GSJ12).
Furthermore, when the spheroid is taken alone, the fits of GSJ12 give
an effective radius of 3.3 kpc, placing it in excellent agreement with the loci
of elliptical galaxies, not bulges, in the mass-size relation.  The similarity
of the spheroid to elliptical galaxies is underscored by its spectacular
globular cluster (GC) population, since the specific frequency of GCs ($S_N$,
which is a measure of the number of GCs per unit galaxy
luminosity in the $V$-band) is discrepant with \textit{any} values seen for disk
galaxies regardless of whether the spheroid is considered separately, but is in
good agreement with values seen for elliptical galaxies \citep{rz04,maybhate}. 
Meanwhile, GSJ12 demonstrate in their Fig.~12 that the Sombrero
is an outlier in several galaxy scaling relations versus mass.  However, when
the disk component from the three-component fits is considered alone, the
Sombrero disk appears reasonably typical of spiral galaxies.
In summary, we are left with a situation where the spheroid and disk,
considered separately, seem fairly typical of elliptical and spiral galaxies
respectively.  However, taken together, the Sombrero as a whole does not appear
typical of either spirals, lenticulars, or ellipticals.  In this context,
further insight into the nature of the spheroid is key to resolve the formation
history of the Sombrero, with the goal of discriminating whether it bears a
closer resemblance to multi-component halos of Milky Way-mass disk galaxies or
massive elliptical galaxies.

The remainder of this study is organized as follows: In the next section, we
describe our observations and data reduction.  Our technique for deriving a
photometric MDF is explained Sect.~\ref{anasect}, and the resulting MDFs and
radial density profiles are presented Sect.~\ref{resultsect}.  In
Sect.~\ref{discusssect} we compare our results with other nearby massive 
galaxies and predicted formation scenarios for the Sombrero, and in the final
section we summarize our results.

\section{Data} \label{obssect}
\subsection{Observations}

The observations we analyze consist of simultaneous \textit{Hubble Space
  Telescope} imaging of two fields located approximately along the minor axis
of the Sombrero, obtained in coordinated parallel mode (GO-14175, PI: P.~Goudfrooij).
Specifically, a WFC3/UVIS field located 
at 15.6 kpc and an ACS/WFC field at 32.5 kpc from the center of M\,104 were
imaged between 18 May and 14 Jun 2016 in the {\it F606W} and {\it F814W} filters of
each instrument by obtaining 52 individual exposures (32 in {\it F606W} and 20 in
{\it F814W}) per instrument.   
The individual exposures were divided into three categories, with 5 (2) short
exposures of 250\,--\,310s each, 15 (6) medium exposures of 600\,--\,700s and 12 (12)
long exposures of 1315\,--\,1400s in {\it F606W} ({\it F814W}), with individual exposure
times varying slightly depending on instrument.  The locations of the two
fields and their total exposure times are summarized in Table \ref{obstab}.

\begin{deluxetable*}{lccccccccc}
\tablecaption{Sombrero Galaxy Minor Axis Halo Fields \label{obstab}}
\tablehead{
\colhead{Instrument} & \colhead{RA (J2000)} & \colhead{Dec (J2000)} & \colhead{L} & \colhead{B} & \colhead{$D_{\rm{M104}}$} & \colhead{$D_{\rm{M104}}$\tablenotemark{a}} & \colhead{$E(B-V)$\tablenotemark{b}} & \colhead{$t(F606W)$} & \colhead{$t(F814W)$} \\ 
\colhead{} &  \colhead{$^\circ$} & \colhead{$^\circ$} &  \colhead{$^\circ$} & \colhead{$^\circ$} & \colhead{arcmin} & \colhead{kpc} & \colhead{mag} & \colhead{s} & \colhead{s}  
}
%\colnumbers
\startdata
ACS/WFC   & 189.9604 & $-$11.4317 & 298.3805 & 51.3375 & 11.69 & 32.5 & 0.0385 & 26150 & 19960 \\
WFC3/UVIS & 189.9862 & $-$11.5301 & 298.4322 & 51.2412 &  5.62 & 15.6 & 0.0323 & 28850 & 21580 \\
\enddata
%\tablecomments{Comments}
\tablenotetext{a}{Assuming a distance of 9.55 Mpc for M104 from \citet{sombrero_trgb}}
\tablenotetext{b}{\citet{sf11}}
\end{deluxetable*}

\begin{figure*}
%\gridline{\fig{FOV_3panels.pdf}{0.9\textwidth}{}}      	 
\gridline{\fig{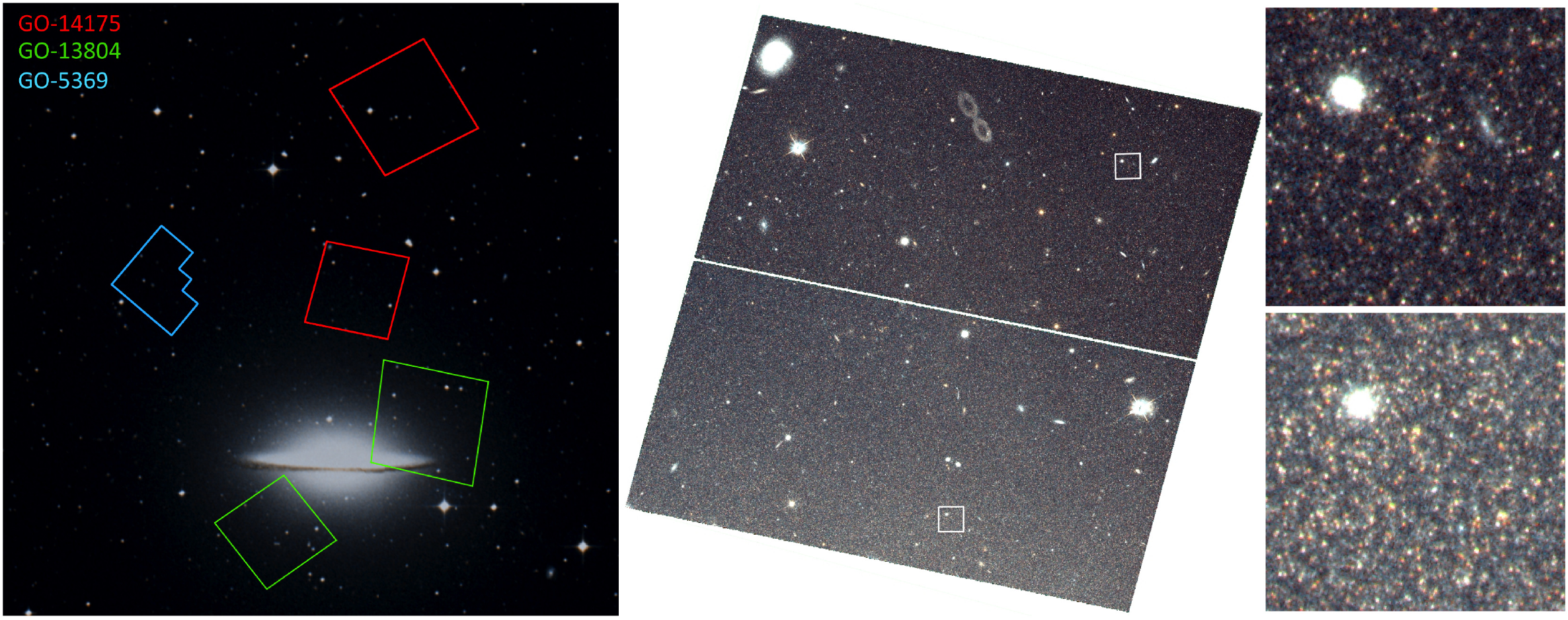}{0.95\textwidth}{}}
\caption{\textbf{Left: }Footprints of recent \textit{HST} imaging of M104.  The
  two fields analyzed here are shown in red, the shallower fields observed in
  GO-13804, including the field used to measure a TRGB distance by
  \citet{sombrero_trgb}, are shown in green, and the shallow WFPC2 field used
  to infer an MDF by \citet{sombrero_wfpc2} is shown in cyan.  \textbf{Center:
  }A three-color image of our UVIS field at $\sim$16 kpc along the minor axis,
  constructed by using the deep drizzled {\it F606W} and {\it F814W} images for the
  blue and red channels respectively, and the average of the two for the green
  channel.  The stellar density gradient with distance from M104 is immediately
  apparent.  \textbf{Right: }Zoomed in color images of the two fields indicated
  by white boxes in the center panel to highlight the stellar densities at
  different locations.  Each image is 10$\arcsec$ on a side and the top and
  bottom images contain the globular clusters RZ3174 and RZ3026 respectively
  \citep[][and references therein]{dowell} in the upper left corner.  In all
  panels, north is up and east is to the left.  \label{fovfig}} 
\end{figure*}

In the left panel of Fig.~\ref{fovfig}, we show the location of the ACS and
UVIS fields, overplotted in red on a DSS image of M104.  We also show in green
the shallower ACS and UVIS fields from GO-13804, a subsection of which was used
to measure a TRGB distance to M104 by \citet{sombrero_trgb}, as well as the
WFPC2 field used to obtain an MDF by \citet{sombrero_wfpc2} in blue.  

\subsection{Preprocessing and Photometry \label{photsect}}

To create a deep stacked distortion-corrected reference image for each filter, we started from the \texttt{.flc} files provided by the \texttt{calacs} or \texttt{calwfc3} pipeline. The \texttt{.flc} files constitute the bias-corrected, dark-subtracted, and flat-fielded images, and include corrections for charge transfer inefficiency. First, we aligned all the \texttt{.flc} files for a given filter to each other, adopting the first exposure as the reference frame, using the software TWEAKREG, part of the {\it Drizzlepac} package \citep{gonz+12}. The transformations between the individual images are based on the fitted centroid of hundreds of stars on each image and the solution was refined through successive iterations, providing an alignment of the individual images to better than 0.04\,--\,0.05 pixels. Then, to remove the geometric distortion, correct for sky background variations, flag and reject bad pixels and cosmic-rays, and combine all the different individual exposures, we used the software {\it AstroDrizzle}, also part of the {\it Drizzlepac} package. The final stacked images were generated at the native resolution of WFC3/UVIS and ACS/WFC (i.e., 0\farcs040 pixel$^{-1}$ and 0\farcs049 pixel$^{-1}$, respectively).

With a deep, drizzled, distortion-corrected reference image in hand for each field, 
subsequent preprocessing and photometry of individual science images for each field was performed using version 2.0 of the publicly available \texttt{Dolphot} package\footnote{\url{http://americano.dolphinsim.com/dolphot/}} \citep{dolphin}.  
Preprocessing was carried out according to the recommendations of each
(instrument-specific) \texttt{Dolphot} manual, including masking bad pixels, separating each science image into individual chips for photometry, and calculating the sky background\footnote{We use the high-resolution \texttt{step}=4 value for generating sky frames.}.  

\texttt{Dolphot} uses customized PSFs tailored to each filter of each \textit{HST} instrument to perform iterative PSF photometry simultaneously across multiple science images which are aligned positionally to the deep distortion-corrected reference image.  
After experimentation with numerous reduction strategies, we decided to perform a single \texttt{Dolphot} run on all images in both filters for each chip of each instrument.  By comparing completeness limits, photometric errors and color and magnitude offset (bias) as a function of color-magnitude location across test runs, we found that \texttt{Dolphot} runs which were separated by either filter and/or exposure length and then matched \textit{a posteriori} 
yielded results which were similar or inferior to performing a single run per detector chip on all 52 images.

Many optional parameters governing how image alignment, PSF fitting and sky subtraction are performed may be altered within \texttt{Dolphot}, and in cases of severe stellar crowding, modifying these parameters can result in deeper, more complete photometric catalogs \citep[e.g.][]{williamsphat,cohenbgc}.  Therefore, we adopt the parameters used by \citet{williamsphat} given the crowded nature of our fields \citep[e.g.][]{dong_galcen_var,conroydolphot}.  This set of \texttt{Dolphot} parameters includes setting \texttt{Force1}=1, which effectively trades away the ability to use the Object Type parameter for star-galaxy discrimination in exchange for deeper, more complete photometry.  Therefore, this choice requires judicious use of several photometric quality diagnostics output by \texttt{Dolphot} to cull non-stellar sources (including image artifacts, background galaxies and globular clusters) from our catalogs.  The photometric quality diagnostics included in the raw catalogs output by \texttt{Dolphot} include: The shape parameters \texttt{sharp} and \texttt{round}, signal to noise (S/N), the \texttt{crowd} parameter indicating how much brighter each star would have been (in magnitudes) had it not been simultaneously fit with its neighbors, and the $\chi^{2}$ of the PSF fit.  For each star, these parameters are given per image, per filter (combined values for all images of a given filter) and per star (combined values for all images in all filters).

The measurement of photometric metallicities requires accurate color information as well as accurate magnitude information, so we apply photometric quality cuts to the per-filter values in order to require high-quality imaging in both bands.  
Our per-filter photometric quality cuts are based on two primary criteria: The
first is examination of the recovered values for input artificial stars over a
range of position (i.e.,~projected stellar density), color and magnitude (see
below), under the hypothesis that the loci in (recovered) parameter space which
are devoid of artificial stars should be occupied only by non-stellar or
spurious sources in the raw observed catalogs \citep[e.g.][]{cohenbgc}.  The
second criterion is examination of observed sources passing and failing a
proposed set of quality cuts in both position-color-magnitude space as well as
visual inspection of accepted and rejected sources in the deep drizzled
reference images to ensure that artifacts such as diffraction spikes are
completely eliminated.  Ultimately, sources retained in our final catalog were
required to have S/N~$\geq$~5, $\texttt{crowd} \leq 0.5$, $\chi^{2} \leq 2$,
$\vert$\texttt{sharp}$\vert \leq 0.3$ and a photometric quality flag $\leq 2$ for
each of the two filters.   
In addition, we found that contamination by compact background galaxies was
drastically reduced using a cut on $\vert$\texttt{sharp}$\vert$ versus magnitude
in each filter by fitting a hyperbolic equation of the form
$\vert$\texttt{sharp}$\vert$$<$$ A + B exp (m-C)$ 
\citep{m101_sharpcut,durrell_m81halo}.  Here, m corresponds to the magnitude in
each filter, and we solve for the coefficients A, B and C using non-linear least
squares fit to the 99.5\% envelope (calculated in 0.1 mag bins) of recovered
$\vert$\texttt{sharp}$\vert$ for the artificial stars.

\begin{figure*}
\gridline{\fig{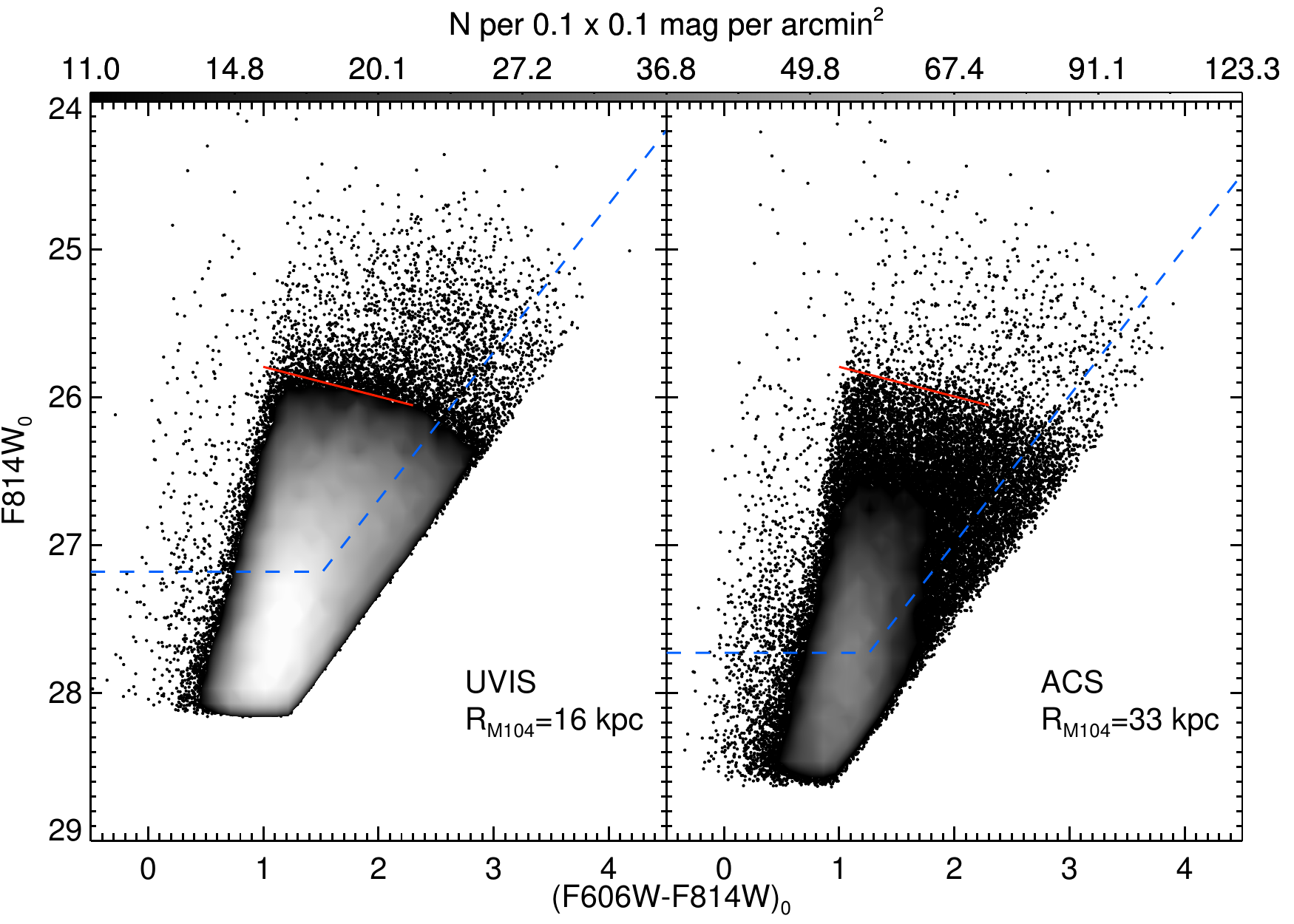}{0.65\textwidth}{}}
\gridline{\fig{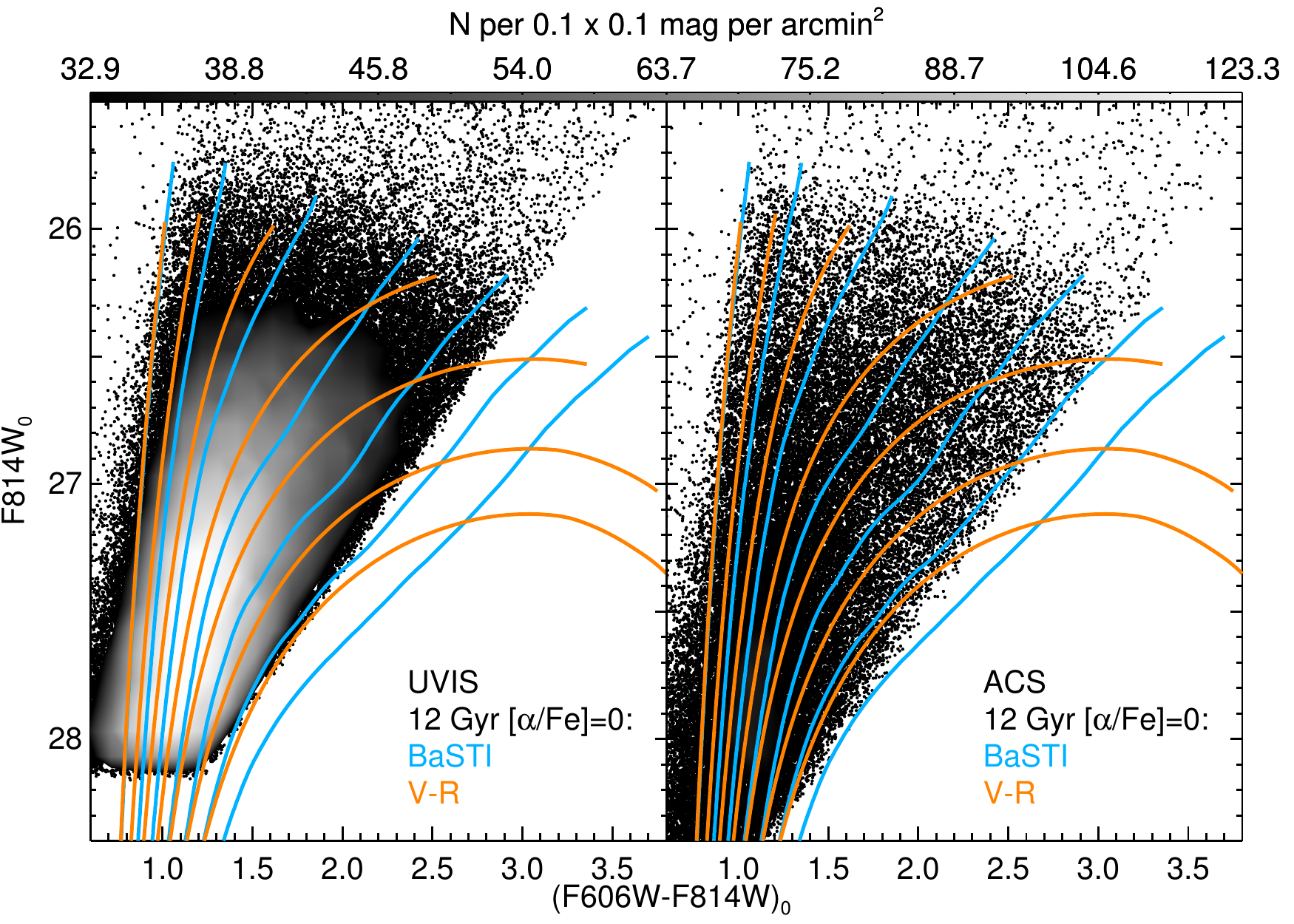}{0.65\textwidth}{}}
\caption{\textbf{Top:} CMDs of the UVIS (left) and ACS (right) target fields,
  where the color scale indicating density according to the colorbars has been
  held fixed between the two fields to enable a direct comparison.  The 50\%
  completeness limits are indicated by a dashed blue line, and the TRGB measured
  by \citet{sombrero_trgb} assuming the \citet{rizzi_trgb} calibration is shown
  as a solid red line.  Photometry has been corrected for foreground extinction
  \citep{sf11} and UVIS magnitudes have been transformed to the ACS/WFC
  photometric system \citep{jang}.  \textbf{Bottom:} Same, but zoomed in on the
  Sombrero RGB with 12 Gyr solar-scaled BaSTI (cyan) and Victoria-Regina
  (orange) isochrones with [Z/H]=[$-$2.3,$-$1.0,$-$0.52,$-$0.20,0,0.18,0.31]
  overplotted assuming $(m-M)_{0} = -29.90$ \citep{sombrero_trgb}. \label{cmdfig}} 
\end{figure*}

The magnitudes output by \texttt{Dolphot} are calibrated to the Vegamag system
using the encircled energy corrections and photometric zeropoints
from \citet{bohlin} for ACS/WFC and from \citet{wfc3_ee,wfc3_isr} for WFC3/UVIS. 
For each field, the photometric catalogs were corrected for
foreground extinction using the \citet{sf11} recalibration of the 
\citet{sfd} reddening maps and the extinction coefficients given in
\citet[][their table A1]{cv14}.  These maps reveal low foreground extinction of
$E(B-V)<0.05$ for all of the sightlines analyzed here.  All of the magnitudes
and colors we report are in the ACS/WFC Vegamag system, corrected for foreground
extinction.  For the UVIS field, foreground-extinction-corrected magnitudes were
transformed to the ACS/WFC photometric system using the empirical relations of
\citet{jang}.  We note that the uncertainties on these transformations are
$\sim$0.01 mag, negligible compared to our photometric errors (ascertained via
artificial star tests) of $\sim$0.1 mag even at the metal-poor TRGB. 
The CMDs of each of the two target fields are shown in Fig.~\ref{cmdfig}.  

\begin{figure*}
\gridline{\fig{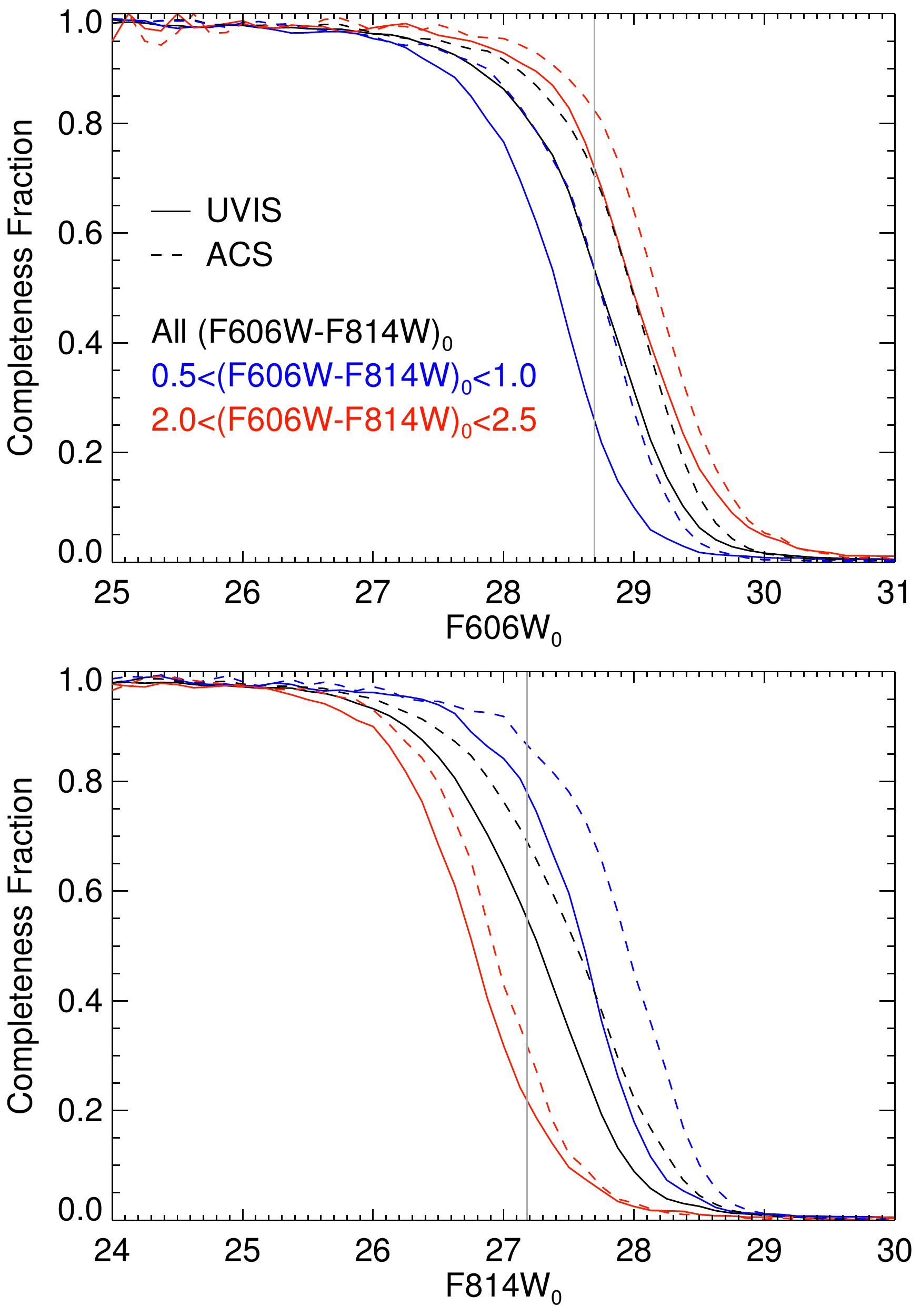}{0.3\textwidth}{}
	  \fig{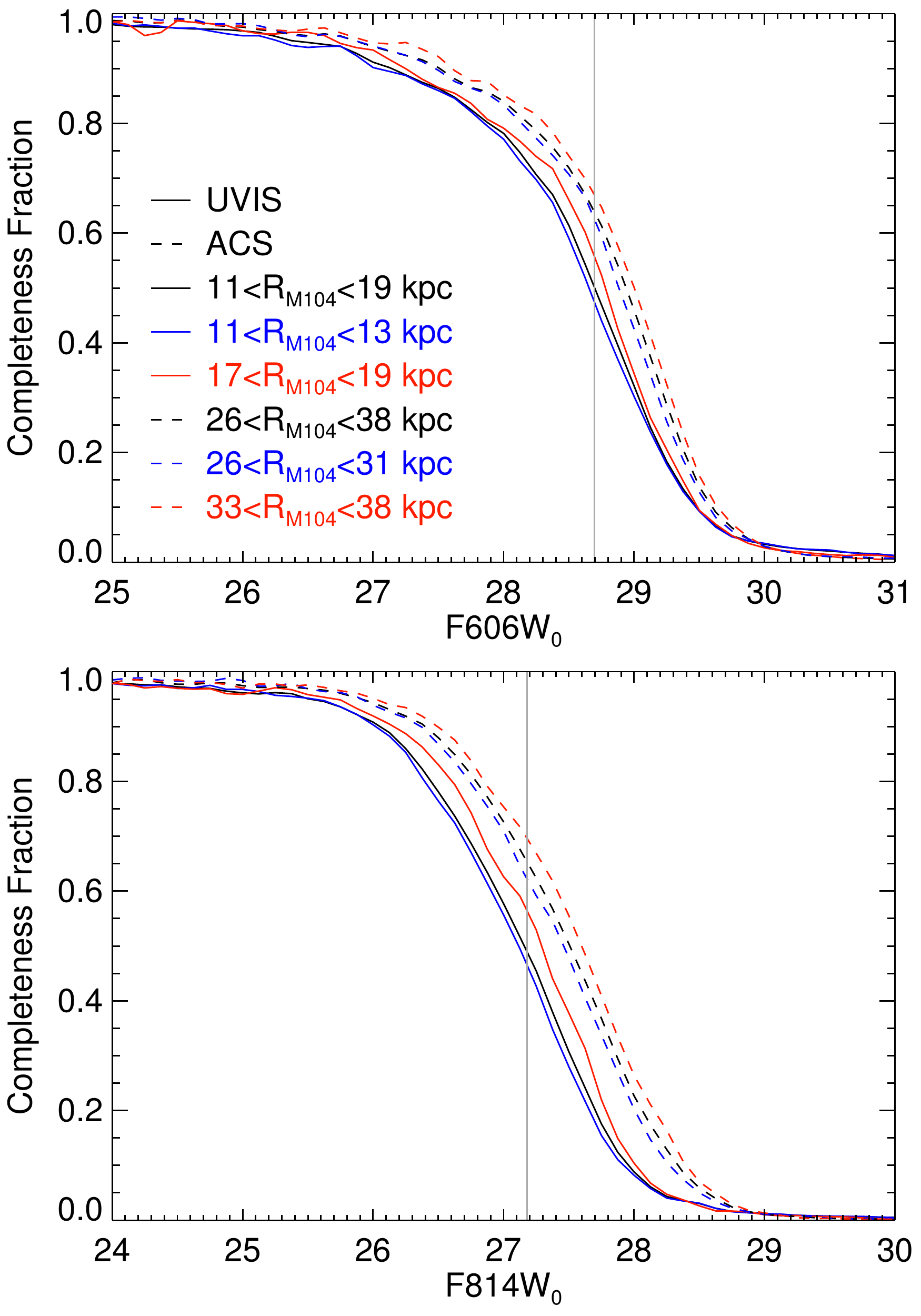}{0.3\textwidth}{}
	  \fig{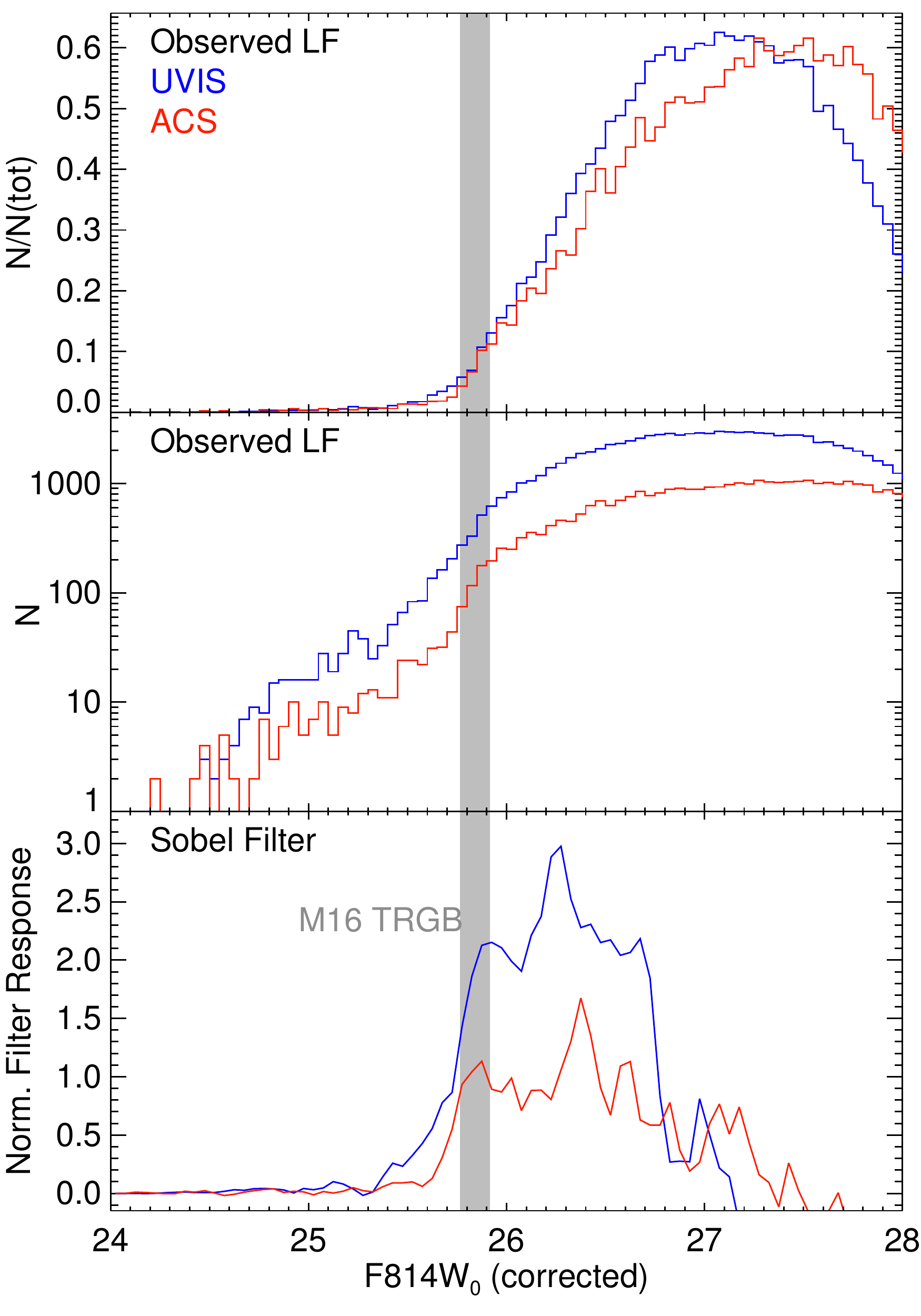}{0.3\textwidth}{}}
\caption{\textbf{Left:} Completeness versus magnitude in the {\it F606W} (upper)
  and {\it F814W}) (lower) filters, where solid lines correspond to the UVIS field and dashed lines correspond to the ACS field.  Completeness curves are color-coded by $(F606W-F814W)_{0}$ color as indicated in the upper panel to highlight the
  dependence of completeness on color at fixed magnitude.  The vertical grey
  line indicates the UVIS 50\% integrated completeness limit adopted for our
  analysis.  \textbf{Center:} Same, but illustrating the dependence of
  completeness on stellar density for each field using two non-neighboring
  radial bins.  \textbf{Right:} Observed luminosity functions for our two
  target fields using {\it F814W}$_{0}$ magnitudes corrected for the color dependence
  of the TRGB \citep{rizzi_trgb} and colors $1 \leq (F606-F814W)_{0} \leq 2.3$.  The upper two panels show the observed LFs on linear normalized and logarithmic y-axis scales to demonstrate the similarity between the two fields in the vicinity of the TRGB, and the bottom panel shows the Sobel filter response.  The brightest Sobel filter peak in both fields is in good agreement with the TRGB magnitude reported by \citet{sombrero_trgb}, indicated by the shaded grey region (including both statistical and systematic errors).  The slight broadening of the UVIS LF near the TRGB as well as its brighter normalized LF peak at the faint end are due to lower completeness caused by increased crowding. \label{completelffig}}
\end{figure*}

Artificial star tests were performed, also using \texttt{Dolphot}, to quantify
incompleteness and photometric errors and offsets in our catalogs.  In order to
reliably sample these quantities over the full range of observed values,
hundreds of thousands of artificial stars were generated for each field.  To
ensure realistic consideration of the effects of blending and crowding as well
as the necessary spatial and color-magnitude coverage, artificial stars were
assigned {\it F814W} magnitudes drawn from an exponential luminosity function
\citep[e.g.][]{natafrgbb}, noting that inserting artificial stars down to more
than 3 mag faintward of our 50\% completeness limits (${\it F814W} \sim 31$)
allows us to quantify the effects of blending on our observed catalog.
Artificial stars were assigned colors to densely sample the color-magnitude
distribution of stars with metallicities ranging from $-$2.5 to 0.5,  
including additional scatter of $> 0.1$ mag in color and magnitude to account for
uncertainties in distance, reddening and photometric zeropoints plus
model-to-model variations in isochrone predictions.   
The artificial stars were assigned an input spatial distribution corresponding
to a power law density profile with power law exponent $\sim -2$ based on the
surface brightness profile of GSJ12 and the globular cluster density profile of
\citet{moretti_gcs}. 
Artificial stars were photometered one at a time so that they are susceptible to
crowding effects from real stars but not from other artificial stars, and were
considered recovered if they passed all of the quality cuts described above.

In Fig.~\ref{completelffig} we plot completeness versus magnitude for both
filters of both instruments, illustrating the strong dependence on color (left
panel) and a more modest dependence on projected stellar density, varying with
distance from M104 (middle panel).   These trends illustrate the necessity to
use a large number of artificial stars to fully map incompleteness as a
function of all three of these observables (color, magnitude and projected
density).  In addition to incompleteness, photometric errors and bias must be
similarly mapped over the entire CMD region of interest in order to translate
our observations to the true MDFs from which they are generated.

\section{Analysis \label{anasect}}
\subsection{Strategy \label{strategysect}}

\begin{figure*}
\gridline{\fig{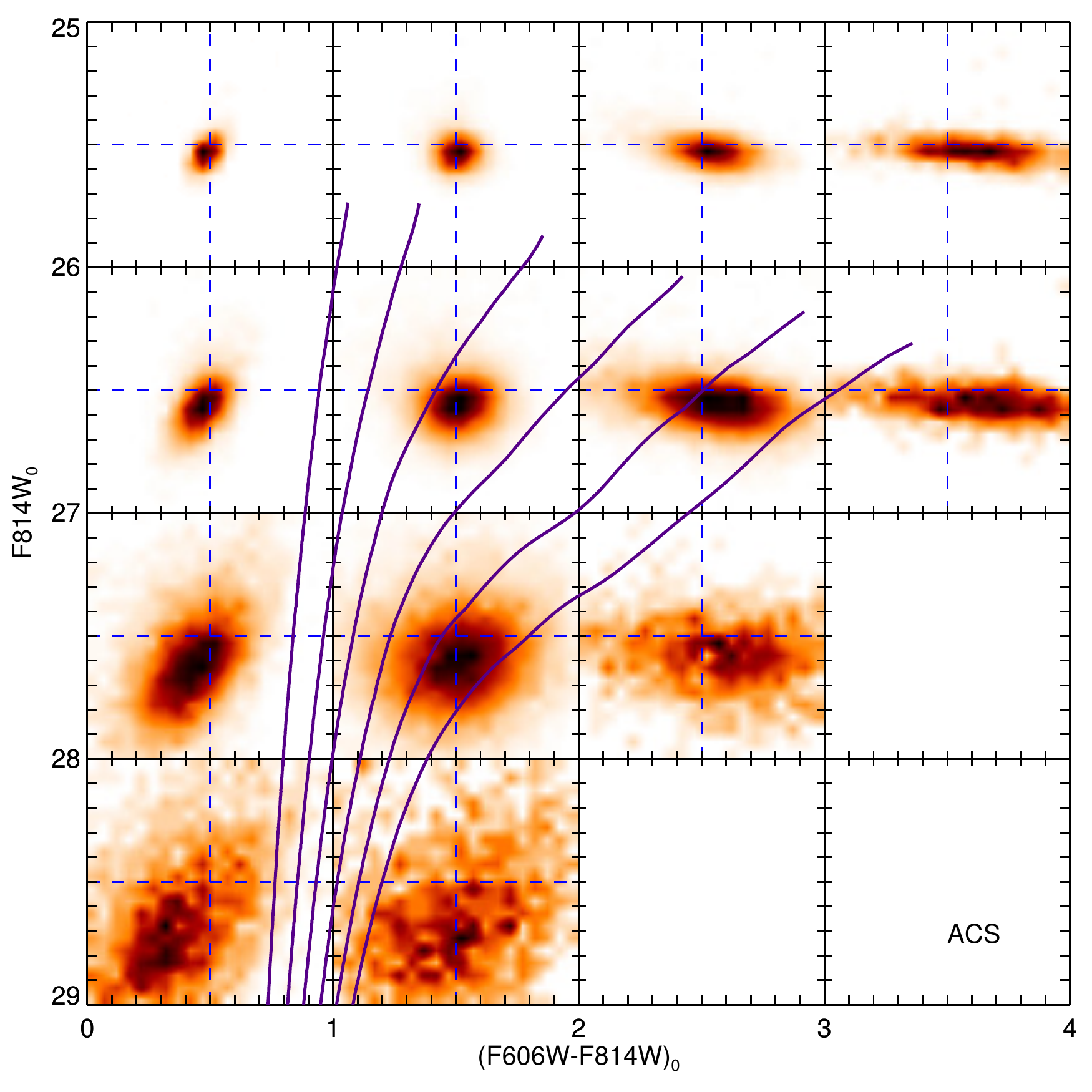}{0.49\textwidth}{}
	  \fig{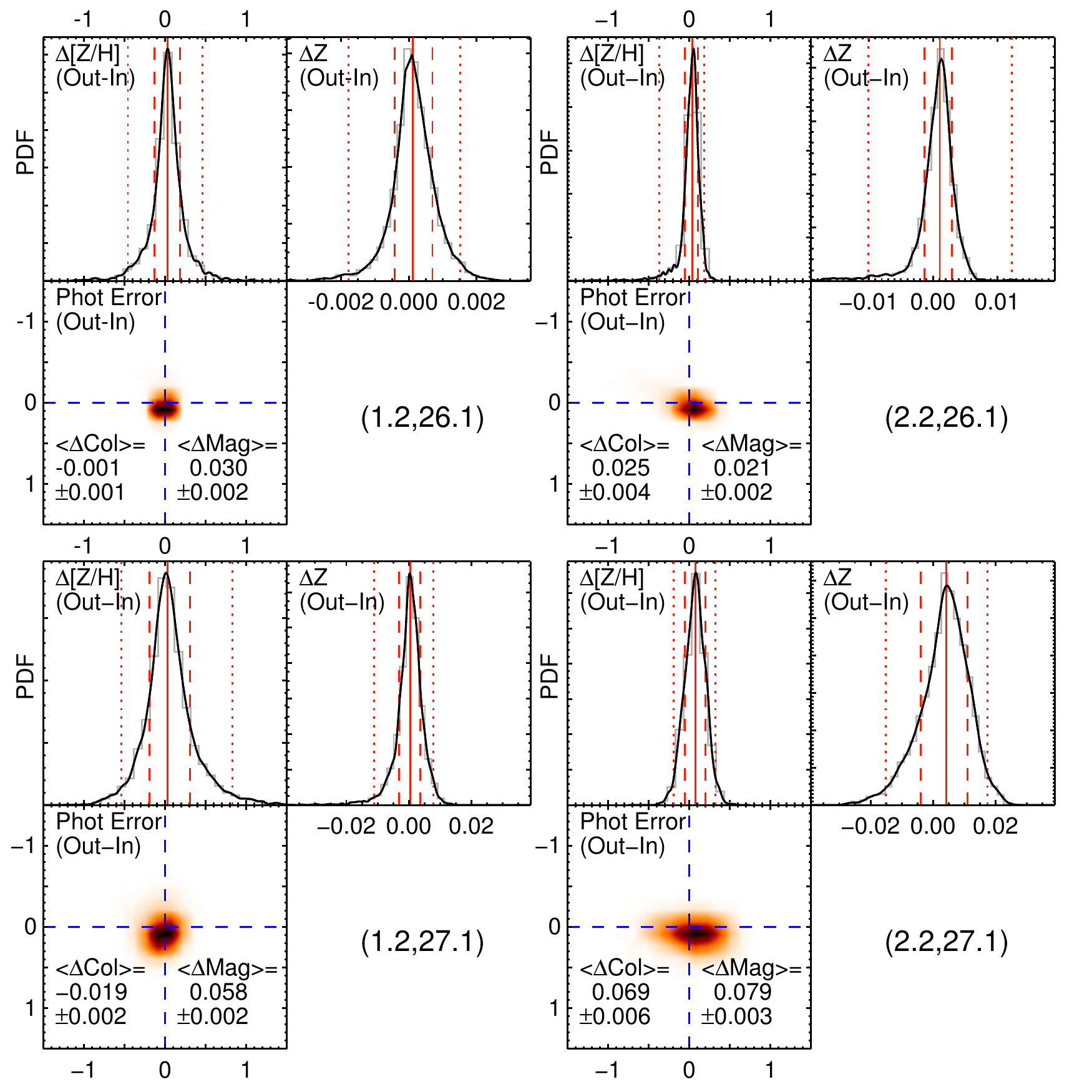}{0.49\textwidth}{}}
\caption{\textbf{Left:} Photometric errors as a function of CMD location for
  the example case of the ACS field, evaluated using artificial star tests.
  For each CMD box, color and magnitude errors are plotted in the sense
  (recovered-input).  Solar-scaled 12 Gyr BaSTI isochrones from
  $-2.3 \leq \mbox{[Z/H]} \leq +0.3$ (only shown through the RGB tip for
  clarity, identically as in Fig.~\ref{cmdfig}) are overplotted in purple.  \textbf{Right:} For four example CMD
  locations (given in the lower right of each set of plots), the lower plot
  shows the photometric error distribution in the same way as the left panel, and the mean bias (offset) in color and magnitude are given, along with their uncertainties.  The upper plots show the difference between output and input
  \textit{metallicity}, also in the sense (recovered-input), as a function of
  global metallicity [Z/H] = log (Z/Z$_{\sun}$) (upper left) and heavy element
  abundance Z (upper right).  The grey histogram displays the raw 
  values, the black line displays the density function obtained using kernel
  density estimation, and the vertical red solid, dashed and dotted lines
  display the median offset, 1$\sigma$ and 2$\sigma$ intervals
  respectively.   \label{kernelfig}} 
\end{figure*}

Using resolved stellar photometry of the upper RGB to measure MDFs
photometrically has the advantage that at high S/N, location on the CMD is
relatively sensitive to changes in metallicity compared to photometric errors.
However, the exact interplay between observational uncertainties and the
resulting uncertainties on photometric metallicity depends on a combination of
stellar parameters (metallicity itself, luminosity, temperature, and to a lesser
extent $\alpha$-enhancement and age) plus observational effects
(i.e.~photometric errors and offsets as a function of color, magnitude and
crowding).  While photometric MDF analyses often assume that photometric errors
are Gaussian, or their influence on metallicity (in terms of either $Z$ or log $Z$)
is Gaussian, there are three ways in which this oversimplification can
quantitatively affect the resulting MDF, particularly at low to moderate S/N, as
is the case for our observations: 
%, illustrated in Fig.~\ref{kernelfig}:

\begin{enumerate}

\item{Color and magnitude errors are almost always correlated, especially for
  crowding-limited imaging.  This correlation generally functions advantageously,
  in that the sense of the correlation serves to scatter stars more parallel to
  the isochrones rather than orthogonally, although the extent to which this is the case depends on CMD location.  This is illustrated in the left
  panel of Fig.~\ref{kernelfig}, where photometric errors from artificial star
  tests are plotted in the sense of recovered-input magnitude in each box,
  analogous to the scattering kernels of \citet{brown_kernels}.  A subset of 12
  Gyr solar-scaled BaSTI isochrones \citep{basti1,basti2} are overplotted in
  purple, increasing in metallicity from left to right.}

\item{At low to moderate S/N, there is a mean offset (bias) between input and
  recovered color and magnitude, and this effect worsens with decreasing S/N and
  increased crowding.  This is seen in the left panel of Fig.~\ref{kernelfig},
  where the observed photometric error distributions are shown for a set of CMD
  locations, each compared to a null offset between input and recovered
  magnitudes indicated by the dotted blue lines.  Furthermore, it is apparent
  that at low S/N, the photometric error distributions become inconsistent with
  a bivariate Gaussian, even once the correlated nature of color and magnitude
  errors is accounted for.} 

\item{The previous two effects often conspire to render the distribution of
  metallicity error non-Gaussian.  This is shown in the right-hand panel of
  Fig.~\ref{kernelfig} for four CMD locations.  For each of the four example CMD
  loci, there are three plots: The lower plot shows the photometric error
  distribution ascertained from artificial star tests (as in the left panel), and the mean spatially-integrated color and magnitude biases and their uncertainties are reported, highlighting that these biases are statistically significant, non-negligible in amplitude, and highly sensitive to CMD location (and also, though not shown here, crowding, as revealed by a comparison between biases in the UVIS and ACS fields at the same CMD locations).   
  The upper two plots corresponding to each CMD location show histograms of recovered-input metallicities (shown
  versus [Z/H] = log Z/Z$_{\odot}$ in the upper left panel and versus Z in the
  upper right panel) based on interpolation between the isochrones.   
  
  While it may not be surprising that some of the resulting distributions of
  metallicity error are non-Gaussian, two subtleties deserve mention: First, the
  CMD locations with relatively Gaussian, symmetric photometric error
  distributions can yield substantially non-Gaussian distributions of metallicity
  error, and vice versa.  Second, whether asymmetric photometric error
  distributions map to asymmetric distributions in either [Z/H] and/or Z depends
  on the \textit{combination} of CMD location and observational effects.  The
  implication is that any assumption on the functional form of either the
  photometric error distribution or the metallicity error distribution cannot
  safely be assumed as valid over the entire sample CMD region.  Therefore, the
  direct use of a large number of artificial stars is required for a proper
  characterization of photometric errors and bias as a function of color,
  magnitude, and projected stellar density.} 

\end{enumerate}

Considering these effects, realistic attempts to recover MDFs and estimate
their uncertainties from low-to-moderate S/N photometry require forward
modeling of CMDs.  This stems from the fact that even when extensive artificial
star tests are available to quantify the aforementioned observational biases,
one cannot \textquotedblleft unscatter\textquotedblright{} observed colors and
magnitudes on a star-by-star basis.  In other words, with only output
photometry available for the observed sample, the metallicity distribution one
obtains by simply interpolating in an isochrone grid on a star-by-star basis
will be sensitive to the assumed input artificial star distribution.  To
alleviate this issue, we take the approach of forward modeling the CMD as a
linear combination of simple stellar populations (SSPs).  In general, this
procedure comes at the cost of sacrificing a strictly continuous metallicity
distribution to instead obtain a piecewise one comprised of various SSPs.
However, in the present case this cost is essentially nullified since our
photometric errors result in 1$\sigma$ metallicity uncertainties of
$\gtrsim$0.15 dex even where the metallicity resolution is highest, and several
times worse over much of the M104 RGB, evident in the right-hand panels of
Fig.~\ref{kernelfig}.    
   
\subsection{Maximum Likelihood Fitting of CMDs \label{sfherrsect}}

We implement forward modeling of observed CMDs for each field using the star
formation history code StarFISH \citep{starfish,starfish2}.
We assume an age of 12 Gyr based on spectroscopy of globular
  clusters in the Sombrero \citep{larsen02,hempel07}.  
The CMDs are assumed to be a linear combination of solar-scaled isochrones, and
in order to investigate the effects of model-to-model variations,  
we include both BaSTI models and Victoria-Regina
\citep[V-R;][]{vrisos} models.\footnote{We also used BaSTI models
  with [$\alpha$/Fe] = +0.4 for comparison purposes, but found that
  the results were indistinguishable when expressed in [Z/H].}      
For each isochrone set, we employ sixteen 12 Gyr isochrones with
$-2.30 \leq \mbox{[Z/H]} \leq +0.31$ spaced such that $\Delta Z/Z \leq 1$, and a
subset of these isochrones are shown in the lower panel of Fig.~\ref{cmdfig}.    
In the presence of large photometric errors, too fine a metallicity sampling
can result in large correlated uncertainties between isochrones which are
photometrically degenerate, so to alleviate this issue while maintaining a
useful metallicity resolution, StarFISH gives the option to lock subsets of
isochrones together, meaning that the amplitudes of isochrones in a locked
subset (i.e.~neighboring each other in metallicity) are constrained to vary
together in lockstep.  After experimenting with various strategies to determine 
which yields the highest quality fits while minimizing correlated errors in the
SSP amplitudes of the solution, we employ nine locked subsets of isochrones.
In addition, we allow the amplitude of the foreground plus background
contamination model (see Sect.~\ref{galcontamsect}) to vary, for a total of ten
SSP amplitudes fit to each CMD.  We assume a Salpeter IMF and a binary fraction
of 50\%, although these choices have a negligible effect on our results since
the stellar mass range which may populate our CMD is quite small, $<$25\%
across the entire sampled metallicity range including post-TRGB evolution, but
$<$1\% at fixed metallicity among RGB stars, which dominate the sample. 

For each field, our CMD fitting is restricted to stars lying within the region
defined by several CMD cuts.  Stars are only included in the CMD fitting if
they meet the following criteria: 
\begin{enumerate}
\item{{\it F814W}$_{0} \leq 27.18$, which is the 50\% spatially integrated {\it
    F814W} 
  completeness limit from the UVIS field (shown in Figs.~\ref{cmdfig} and
  \ref{completelffig}).  For consistency, we apply this same limit to the ACS
  field despite its fainter completeness limit.  While many MDF studies employ
  completeness limits in both filters \citep{harris_3377,sombrero_wfpc2}, we
  found that such a cut sacrificed our ability to meaningfully constrain the
  most metal-rich (solar to super-solar) population, while yielding consistent
  results for the remainder of the (sub-solar metallicity) sample.  Conversely,
  making a cut using only one filter \citep[e.g.][]{peacock_lent_mdf} has the
  advantage of yielding statistically meaningful constraints on the super-solar
  metallicity population provided extensive artificial star tests have been
  used to model the variation of incompleteness, photometric error and bias as
  a function of color, magnitude, and projected stellar density.} 
\item{$(F606W-F814W)_{0} \geq 0.5$.  This cut eliminates CMD regions where
  contamination by background galaxies is fractionally large and a negligible
  fraction ($<$1\%) of even the most metal-poor (Z=0.0001) stars are expected
  based on the artificial star tests and isochrones.} 
\item{{\it F814W}$_{0} \geq 26$.  This cut eliminates candidate RGB and AGB variables
  not represented in the SSP models, likely due to their large amplitudes and
  long periods which we are sampling stochastically.  This issue is discussed
  further in the Appendix.}  
\end{enumerate}

STARFish calculates the best-fit amplitudes evaluated using the Dolphin Poisson
statistic \citep{dolphin_sfh}, and accounts for correlated uncertainties among
the SSPs.  However, because the CMD fitting procedure functions by drawing a
finite number of artificial stars, binning the CMD and the artificial star error
distributions, the total uncertainties we report are the quadrature sum of those
from several individual sources \citep[e.g.][]{hidalgo_sfh,bernard_sfh}.  The
first source of uncertainty we consider in the SSP amplitudes is the mean
$\pm$1$\sigma$ errors reported by STARFish over many repeated runs on each field
using the same parameters (i.~e.~CMD binning) and changing only the random seed.
This is necessary to account for stochastic fluctuations stemming from the fact
that STARFish uses binned cumulative distribution functions to store crowding
and photometric error information over the CMD, and accounts for variations
stemming from a finite number of artificial stars populating each isochrone
according to draws from a prescribed IMF.  Because this mean observed
uncertainty is with respect to the best-fit amplitude in each run, the second
source of error we add in quadrature is the 1$\sigma$ (calculated as the 16th
and 84th percentiles) variation in the mean amplitude for each SSP over all of
the runs in which only the random seed is varied.  Next, we perform another set
of runs on each field, but this time varying both the sizes of the CMD bins as
well as the bin locations for each size in order to account for any biases
caused by a particular binning scheme.  In particular, the bin sizes are varied
by $\pm$80\% from our default value of 0.05 mag CMD pixels binned 2$\times$2 to
as small as 0.06 mag (0.03 mag CMD pixels binned 2$\times$2) and as large as
0.18 mag (0.06 mag CMD pixels binned 3$\times$3), and for each bin size, the bin
starting point is shifted in increments of 0.01 mag in color and magnitude.  
As a result, we also add in quadrature the $\pm$1$\sigma$ variations in the mean
amplitude for each SSP over these bin variations, plus any difference between
the mean amplitude from the original runs and the mean amplitude from the runs
allowing for bin variations.  Of the aforementioned sources of error,
reassuringly, the uncertainties output directly by STARFish dominate the total
error in the majority (65\%) of cases.  However, the uncertainty due to varying
the size and location of the CMD bins which are compared to the observations is
non-negligible, dominating the uncertainty in 28\% of the runs (preferentially
for the more crowded UVIS field) and contributing a median of 28\% to the error
budget.   

\subsection{Contaminants}

When fitting our observed CMDs, we must account for the fact that, even given the CMD cuts mentioned in Sect.~\ref{sfherrsect}, the Sombrero RGB is subject to some non-zero contamination by foreground Milky Way stars and background galaxies with star-like PSFs.  Also, blends and unresolved star clusters may contaminate our catalogs, and we now describe how these potential sources of contamination are quantified.

\subsubsection{Foreground Milky Way Stars \label{mwcontamsect}}

We use the TRILEGAL Galaxy model \citep{trilegal1,trilegal2} to predict the
density of foreground Milky Way stars towards each of our two target fields.  For each field, we sample an area of 0.01 deg$^{2}$ on the sky in each TRILEGAL run and concatenate multiple runs to improve Poisson statistics.  We find that
the predicted foreground contamination does not vary significantly between our two fields due to their proximity on the sky, and the majority of the foreground contribution occurs significantly brightward of
the Sombrero TRGB: For both fields, we find 6.3 foreground contaminants per
arcmin$^{2}$ over the entire CMD, but only 1.6/arcmin$^{2}$ anywhere in the
approximate magnitude range of the Sombrero RGB (${\it F814W}_{0}>25.5$)
\textit{before} applying the incompleteness in the target fields.  Of these,
about 1/4 are white dwarfs or brown dwarfs, so after applying incompleteness in
the target fields as a function of color, magnitude and location (assuming a
homogenous spatial distribution for the contaminants), the predicted number of
\textit{recovered} contaminants with ${\it F814W}_{0}>25.5$ drops to 0.58 (0.63) stars 
arcmin$^{-2}$ for the UVIS (ACS) field.  Given the observed source density of well
over 10$^{3}$ arcmin$^{-2}$ even in the far reaches of the more sparse ACS
field, such a contribution by foreground stars is essentially negligible,
although we include it in our contamination model described here. 

\subsubsection{Background Galaxies \label{galcontamsect}}

\begin{figure*}
\gridline{\fig{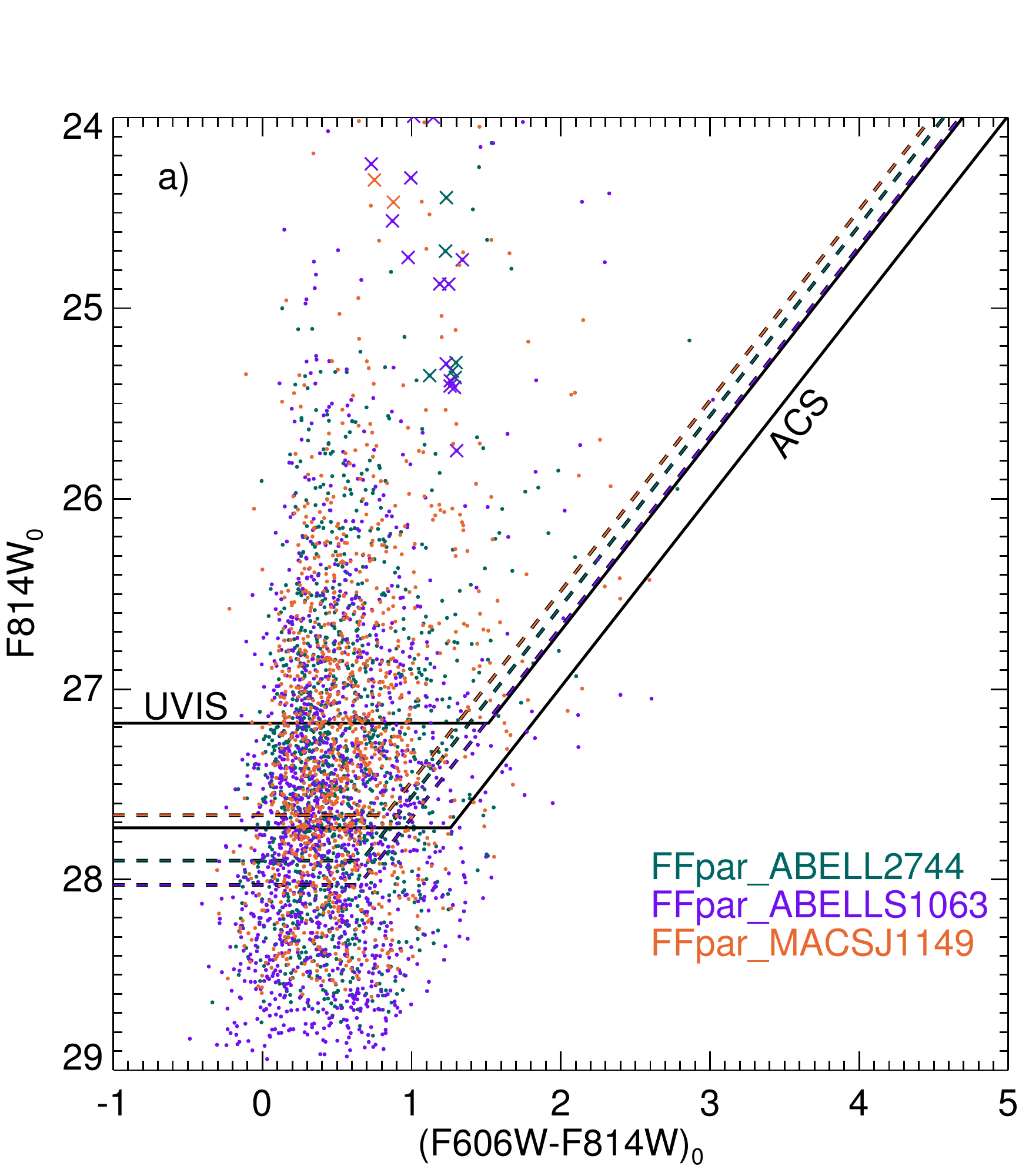}{0.31\textwidth}{}
	  \fig{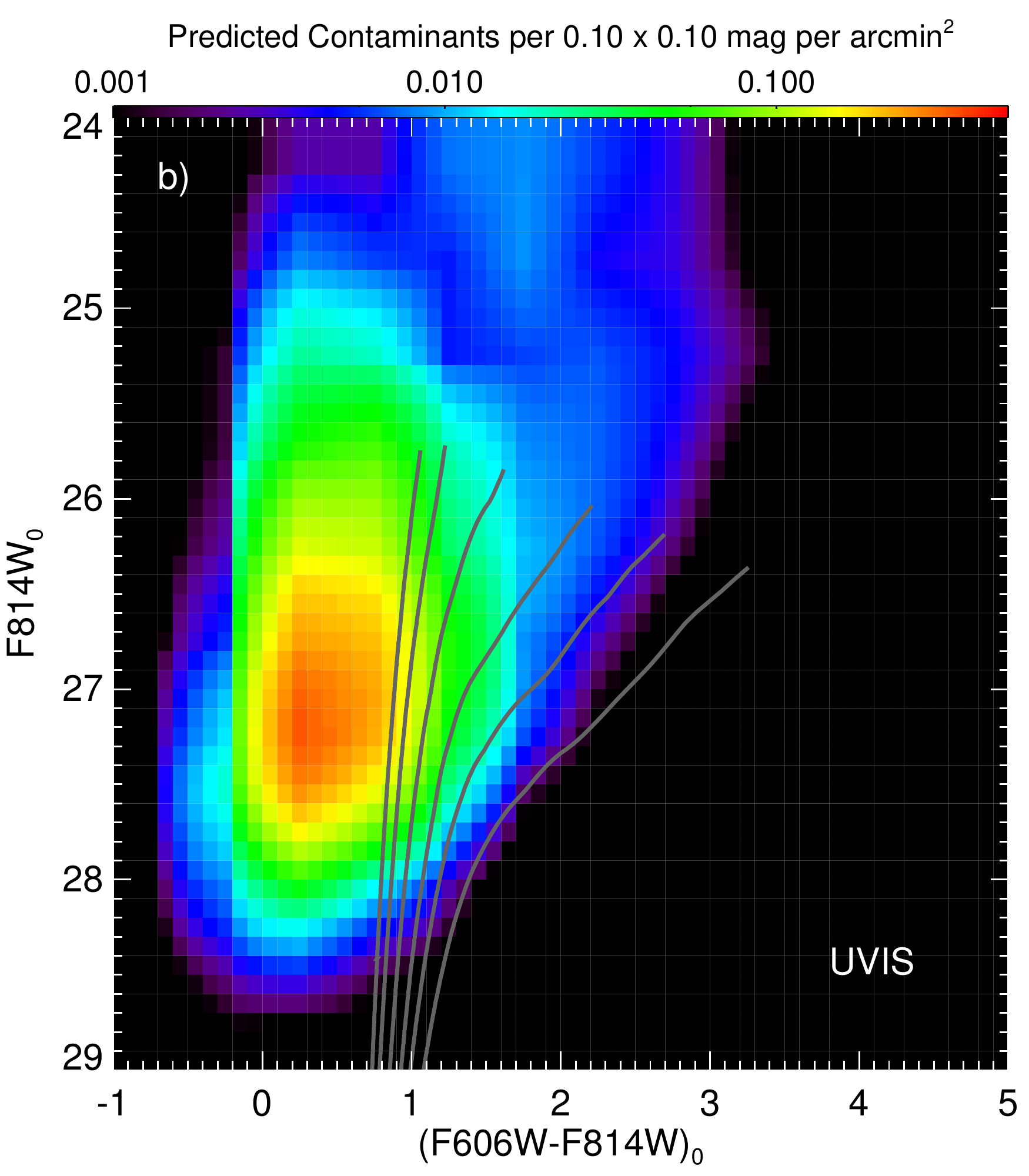}{0.31\textwidth}{}
	  \fig{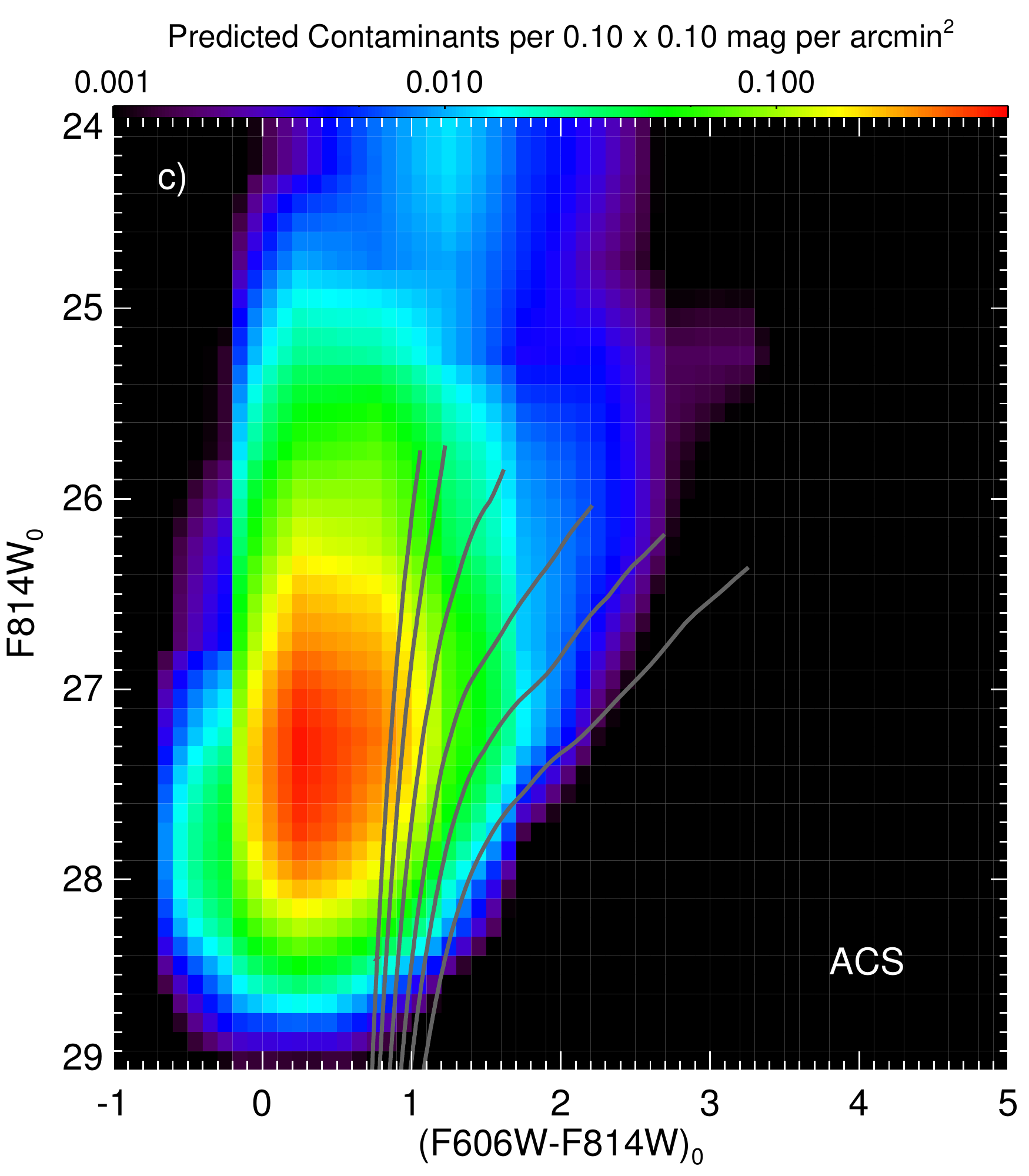}{0.31\textwidth}{}}
\gridline{\fig{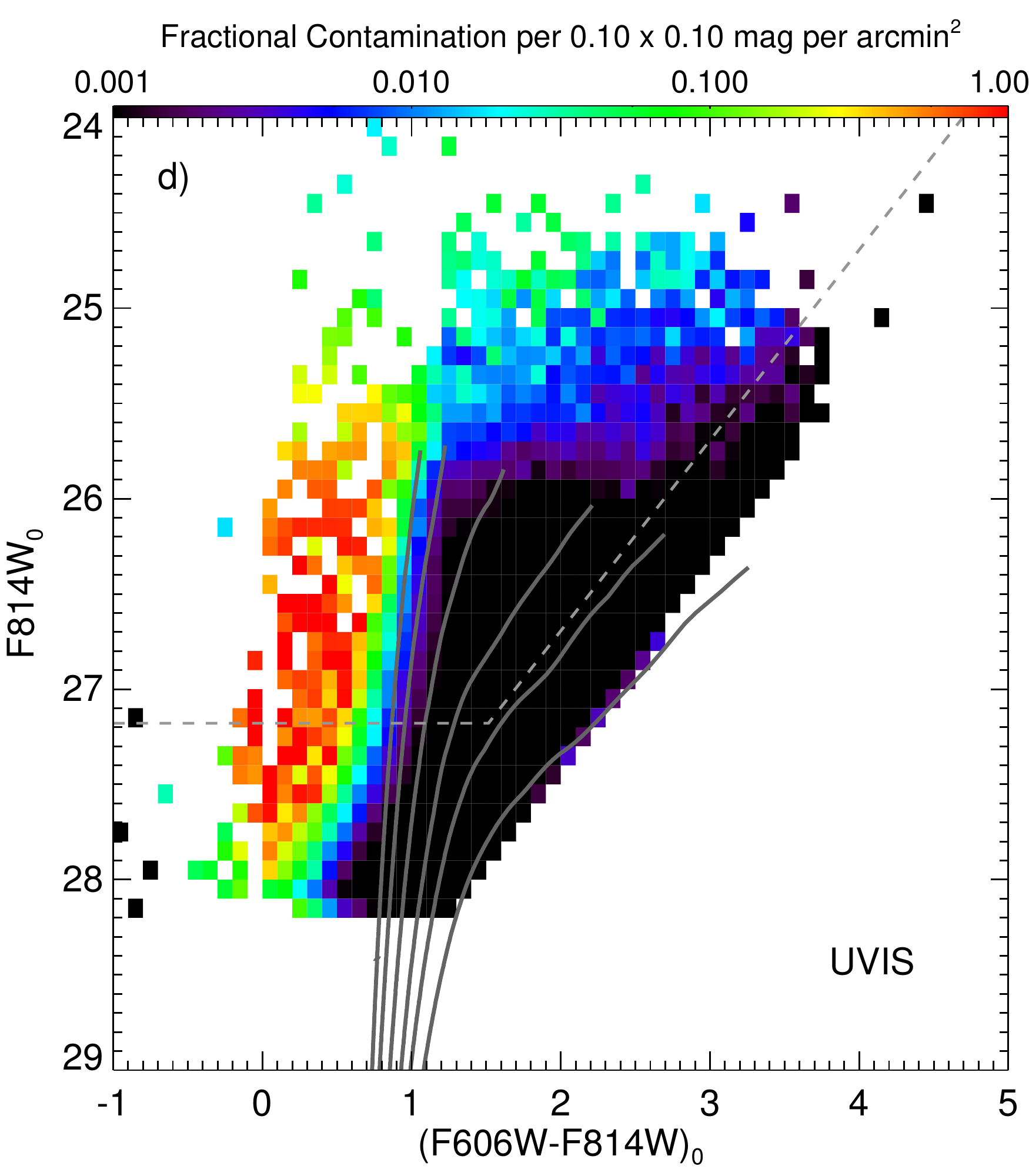}{0.31\textwidth}{}
	  \fig{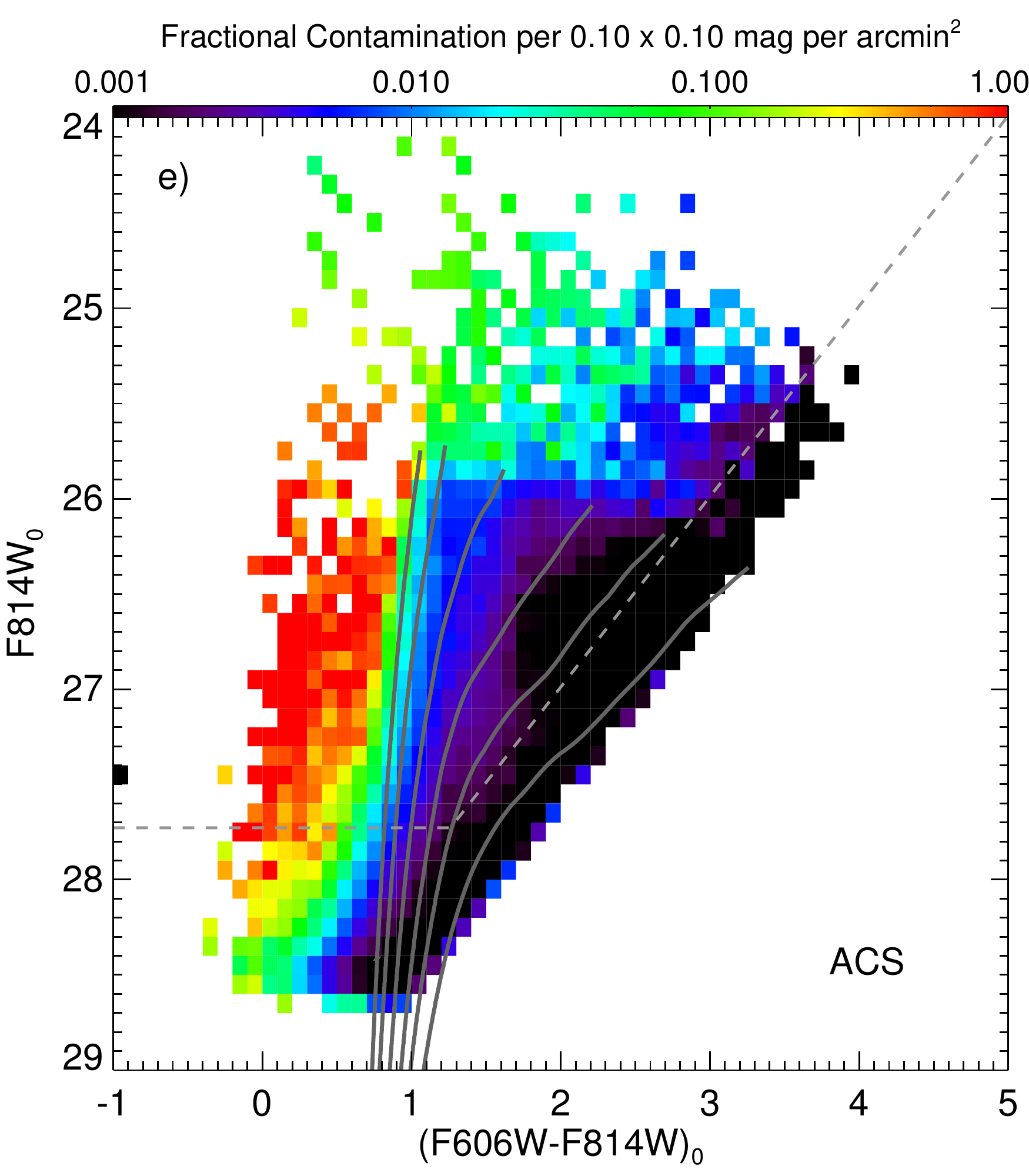}{0.31\textwidth}{}}
\caption{Contamination by background galaxies and foreground Milky Way stars
  towards our target fields.  \textbf{Top Left:} Sources from the blank galaxy
  fields passing all of our photometric quality cuts, color coded by field.
  Probable foreground stars in each blank galaxy field based on the TRILEGAL
  model are shown as crosses and excluded from further analysis.  The 50\%
  completeness limits in each blank galaxy field are shown as dotted lines, and
  the 50\% completeness limits in our ACS and UVIS target fields are shown as
  black solid lines.  \textbf{Top Center and Top Right:} Hess diagram of the
  predicted density of contaminants, including background galaxies plus
  foreground Milky Way stars towards our target fields, after correcting for
  incompleteness in the blank Galaxy fields and applying incompleteness in our
  target fields, for the UVIS and ACS fields respectively.  Curved grey lines
  represent a subset of 12 Gyr solar scaled BaSTI isochrones with metallicity
  $-2.3 \leq \mbox{[Z/H]} \leq +0.3$ assuming $(m-M)_{0} = 29.90$.
  \textbf{Bottom Row:} Contamination \textit{fraction} over the CMD for our
  UVIS (left) and ACS (right) field, shown on a logarithmic color
  scale. \label{galcontamfig}} 
\end{figure*}

We build a model describing the 
distribution of background galaxies predicted using imaging of high Galactic latitude \textquotedblleft blank\textquotedblright{} fields.  The construction of the background galaxy CMD distribution is described in more detail in \citet{m101_sharpcut} and \citet{cohenamiga}, but can be briefly summarized as follows:

First, we searched the \textit{HST} archive for fields far from the Galactic
plane (to minimize the impact of foreground stars) observed in identical
filters and similar (or deeper) exposure times as our science images.  This
search yielded three fields observed as coordinated parallels in the HST
Frontier Fields program, described in more detail in \citet{hff}, and listed in
Table \ref{blankfldtab}.  We then selected a subset of exposures in each of
these fields so as to obtain exposure times similar to our Sombrero imaging,
mimicking the photometric depth of our observations.  We performed \texttt{Dolphot}
photometry and artificial star tests on these images exactly as described in
Sect.~\ref{photsect}, including identical photometric quality cuts.  The
resulting CMD is shown in the upper left panel of Fig.~\ref{galcontamfig},
where sources are color coded by field. The CMD loci of contaminating
background galaxies are essentially identical to what was found in other recent
studies employing a similar strategy \citep{m101_sharpcut,cohenamiga},
consisting of a vertical swath of sources with
$0 \lesssim {\it (F606W-F814W)}_{0} \lesssim 1$ which have a (completeness-corrected)
luminosity function rising towards fainter magnitudes.  To build our
contamination model, we first correct for incompleteness in the blank fields,
and then apply incompleteness towards our Sombrero fields, resulting in the
contamination model shown in the upper right panels of Fig.~\ref{galcontamfig}.
Lastly, in the bottom row of Fig.~\ref{galcontamfig}, we bin the CMD in
0.1$\times$0.1 mag bins to illustrate the \textit{fraction} of sources
predicted to be contaminants as a function of CMD location.  Here it is clear
that the contamination fraction only becomes significant blueward of the
metal-poor Sombrero RGB.   

\begin{deluxetable}{lcclr}
\tablecaption{Blank ACS/WFC Fields Used to Assess Background Galaxy Contamination \label{blankfldtab}}
\tablehead{
\colhead{Field} & \colhead{$t(F606W)$} & \colhead{$t(F814W)$} & \colhead{L} & \colhead{B} \\ 
\colhead{} & \colhead{s} & \colhead{s} & \colhead{$^\circ$} & \colhead{$^\circ$} 
}
%\colnumbers
\startdata
ABELL-2744-HFFPAR & 24223 & 25430 & 9.15  & $-$81.16  \\
MACS-J1149-HFFPAR & 25035 & 23564 & 228.57  & 75.18  \\
ABELLS1063-HFFPAR & 24635 & 23364 & 349.37 & $-$60.02 \\
\enddata
%\tablecomments{Comments}
\end{deluxetable}

\subsubsection{Blends}

The effect of photometric blends on our imaging can be quantified since we
intentionally include artificial stars more than 3 mag faintward of the CMD
region used for our analysis, input with a realistic (exponentially increasing
towards fainter magnitudes) luminosity function.  However, we find that while
photometric errors and bias are non-negligible throughout the observed CMD (see
Fig.~\ref{kernelfig}), cases where much fainter companions are blended with
brighter sources are rare.  Over the entire CMD region sampled by the
artificial stars, extending down to {\it F814W} = 31, less than 1\% of sources
in the CMD region we employ in either the UVIS or ACS fields had an input
magnitude more than 2 mag fainter than its output magnitude.  Similarly, of the
artificial stars with \textit{output} magnitudes falling in the CMD region we
use for our CMD fitting (i.e.~brightward of the UVIS 50\% completeness limit in
{\it F814W}), less than 0.2\% (UVIS) and 0.1\% (ACS) had input magnitudes
faintward of {\it F814W} = 29. 

\subsubsection{Unresolved Star Clusters}

Globular cluster half-light radii range from $\sim$\,1\,--\,10 pc in the MW
\citep[][2010 edition]{harriscat}, the inner regions of M104
\citep{sombrero_gc_phot_3} and other massive ellipticals
\citep{woodley5128,paul_gcs,webbm87gcs,puzia_gcs} as well as young massive
clusters \citep{ymcreview}.  At the distance of the Sombrero, 1 ACS/WFC pix
corresponds to 2.3 pc on a side, so that any clusters in our images should be
distinguishable from stellar sources, which have a well-defined PSF.  For
example, \citet{sombrero_gc_phot_3} were able to measure globular cluster
half-light radii down to $\sim$0.9 pc (corresponding to 0.02$\arcsec$) using
an HST mosaic of the inner region of the Sombrero, and find that there, as in
other galaxies (see their fig.~11), the majority of globulars are larger than
this, with a distribution of half-light radii peaking at $\sim$\,2\,--\,3 pc with a
tail towards larger sizes.  Meanwhile, \citet{sombrero_gc_phot_1} find a
Sombrero globular cluster luminosity function that peaks far brightward of the
Sombrero TRGB, at $V$ = 22.17, and this distribution (see their fig.~3)
implies that even in the more central region of the Sombrero targeted by their
observations, the projected density of globular clusters faintward of their
completeness limit at $V$ = 24.3 is $\sim$\,0.1/arcmin$^{2}$, so the number of
globular clusters coincident with the Sombrero RGB ($V>25.5$) in either of our
fields is expected to be essentially zero.  Furthermore, the radial density
profile of Sombrero globular clusters decreases at increasing projected radii,
so given the \textit{stellar} projected densities of our target fields as well
as the size distribution of globular clusters, the contamination of 
our fields by globular clusters is negligible.

\section{Results \label{resultsect}}  

\subsection{Global Metallicity Distribution Functions \label{sfhwholesect}}

\begin{figure*}
\gridline{\fig{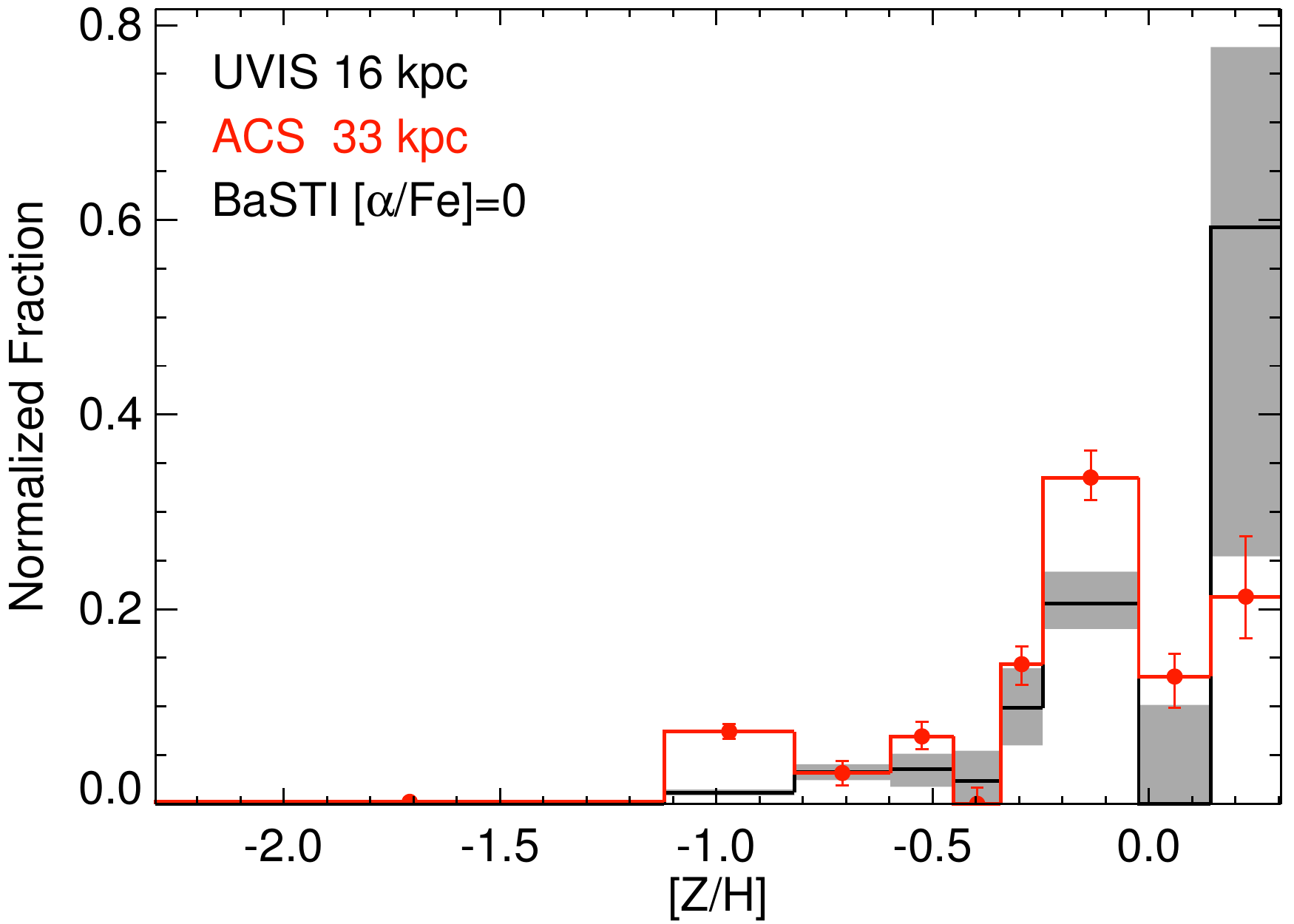}{0.43\textwidth}{}
	  \fig{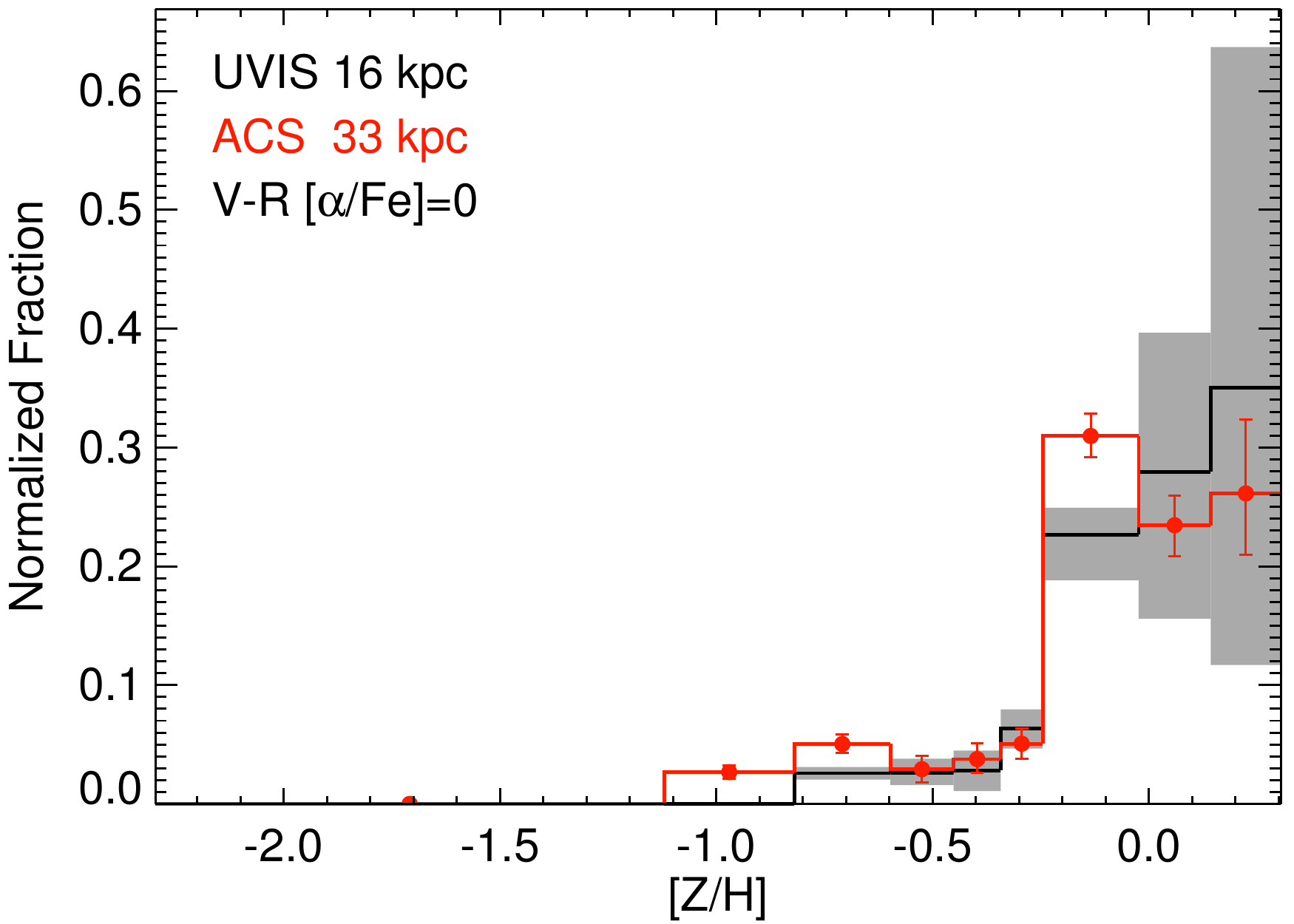}{0.43\textwidth}{}}
\caption{Metallicity distribution functions for the 16 kpc UVIS field, shown in
  black, with total 1$\sigma$ uncertainties (calculated as described in
  Sect.~\ref{sfherrsect}) indicated by the grey shaded region, and for the 33
  kpc ACS field, shown as a red line with errorbars indicating
  uncertainties.  
The left and right panels show results based on scaled solar BaSTI and
Victoria-Regina models respectively. 
\label{mdfwholefig}}
\end{figure*}

Metallicity distributions for the UVIS (16 kpc) and ACS (33 kpc) fields are
shown in Figure \ref{mdfwholefig}, where error bars represent the total 1$\sigma$
uncertainties calculated as described in Sect.~\ref{sfherrsect}.  
Strikingly, both fields are dominated by metal-rich stars, and the most clear
difference between them is a decrease in the relative number of
super-solar-metallicity stars accompanied by an increased
intermediate-metallicity ([Z/H] $\gtrsim -1$) population in the outermost 33 kpc
ACS field.  The fraction of stars with [Z/H] $< -0.6$ increases by a factor of
two to three moving radially outward from the UVIS to ACS fields, and the
fraction of stars with [Z/H] $< -0.8$ increases by a factor of six.  
Focusing on the \textit{relative differences} between the two fields for any set
of model assumptions, both isochrone models are in agreement regarding the
dominance of metal-rich stars and lack of metal-poor stars.  Viewed in terms of
a radial gradient in median metallicity, the models concur on 
a power law gradient of $\delta$[Z/H]/$\delta$(log $R_{\textrm{gal}})\sim -$0.5
(or expressed linearly, $\sim -$0.01 dex/kpc), with a median
$-0.06\leq$\,[Z/H]\,$\leq 0.03$ in the UVIS field and
$-0.19\leq$\,[Z/H]\,$\leq -0.12$ in the more distant ACS field.   
Notably, such a high median metallicity is supported by colors measured from
optical integrated light.  \citet{hargis_gcs} present radial color profiles of
several early-type galaxies, including the Sombrero, and while their profile
extends only to our UVIS field, conversion of their reported colors (and
uncertainties) to metallicity using their eq.~4 gives [Z/H] = $-0.03 \pm 0.18$
over the radial range of the UVIS field, in excellent agreement with our results. 
Perhaps the most striking aspect of the MDFs of the UVIS and ACS fields is the
paucity of metal-poor stars, even in the more distant ACS field where they are
relatively more common: Of the entire population, both models find that stars
with [Z/H] $< -0.8$ constitute less than 10\% there.  Restricting the entire
sample to [Z/H] $< -0.25$ where relative errors are somewhat smaller, stars with
[Z/H] $< -0.8$ still constitute less than half and those with [Z/H] $< -1.2$
have 1$\sigma$ \textit{upper} limits of 1-4\% depending on the adopted isochrone set.

\begin{figure*}
\gridline{\fig{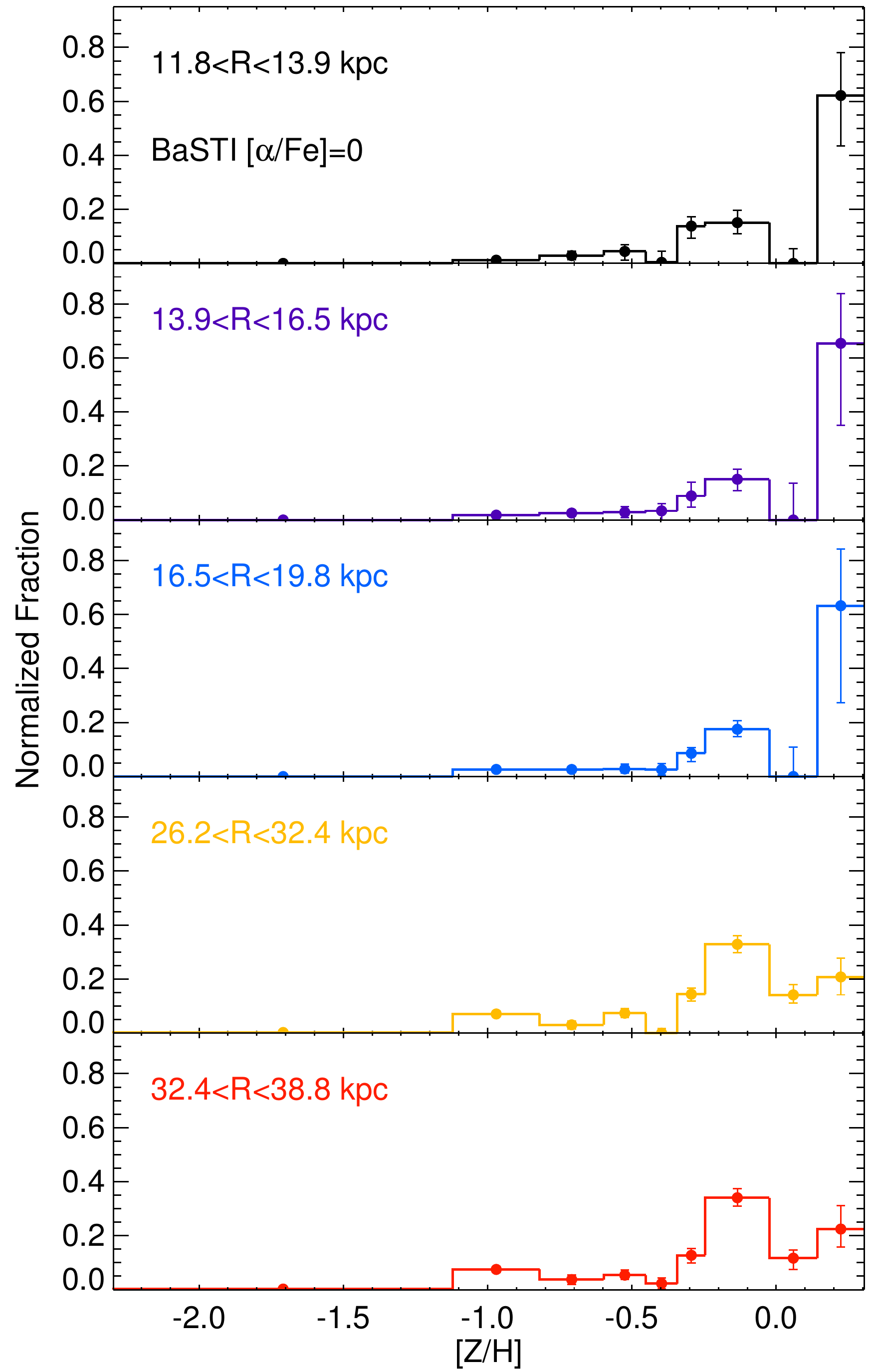}{0.4\textwidth}{}
	  \fig{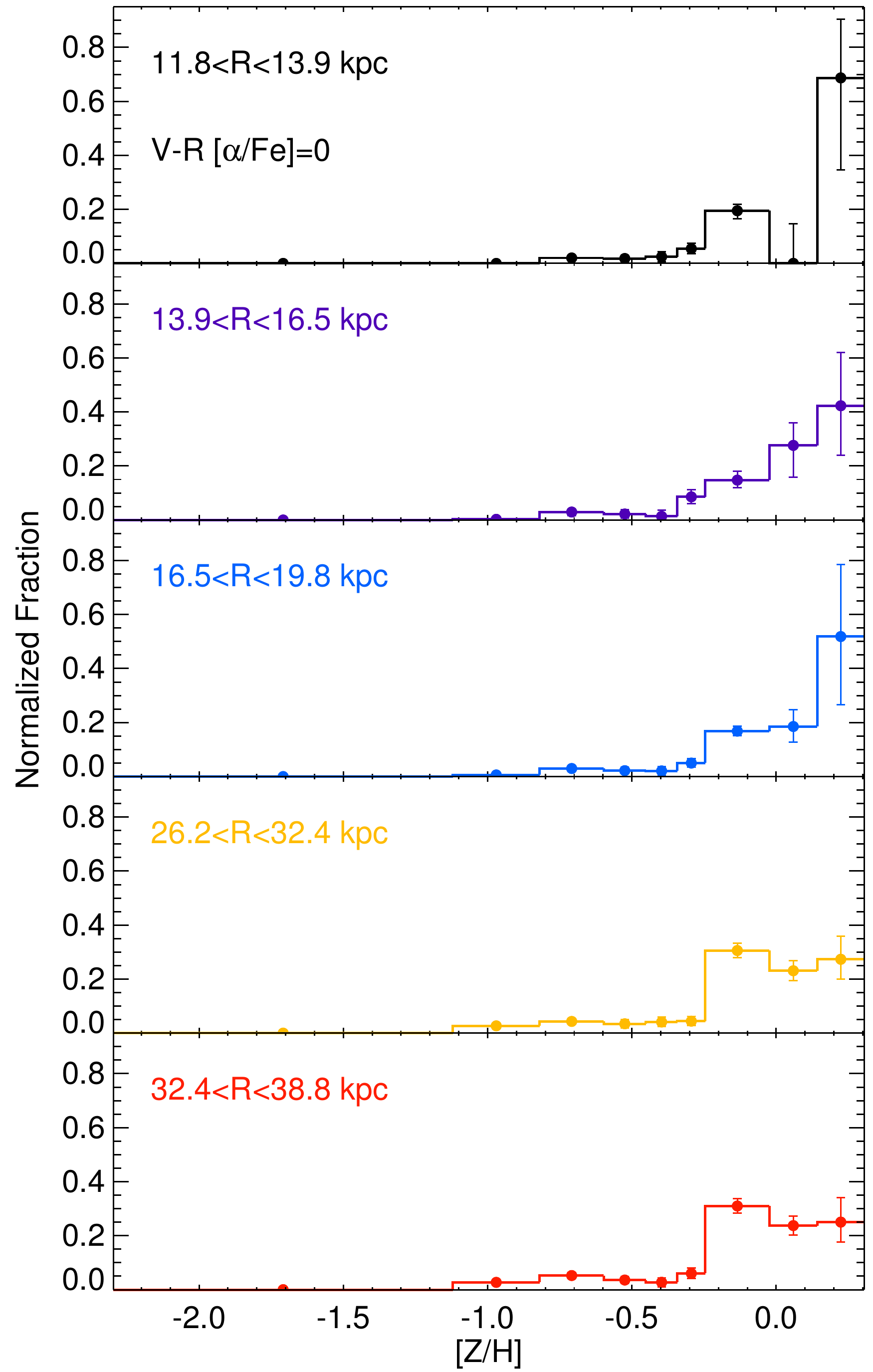}{0.4\textwidth}{}}
\caption{Metallicity distribution functions, normalized to their total.  Each
  column represents a different set of isochrones as in Figure \ref{mdfwholefig},
  but for each assumed isochrone set we now plot separate MDFs for each radial
  bin, indicated in each panel, moving farther from the center of M104 from top
  to bottom in each column. \label{mdfradbinfig}} 
\end{figure*}

To gain further quantitative insight into the dominance of the metal-rich population, and its physical cause via any dependence on projected radius from M 104, we divide our sample spatially by projected radius into bins.  This division is made by keeping the observed number of stars in each bin similar in an attempt to minimize differences in Poissonian uncertainties across the radial bins, and is a necessary compromise between spatial (i.e.~radial) resolution and number statistics.  After testing several radial binning schemes, we employ a total of five radial bins, three in the UVIS field and two in the ACS field. 
We performed our STARFish MDF analysis completely independently on each radial bin, employing only artificial stars located in the corresponding bin.  While this approach comes at the cost of number statistics, it has the advantage of locally sampling the true stellar density in each radial bin, minimizing the effects of (and serving as a check on) a single assumed sample-wide density gradient for the artificial stars.  The resulting MDFs for each radial bin (and the radial bin locations) are shown in Fig.~\ref{mdfradbinfig}, where we see in more detail the decreasing fraction of super-solar-metallicity stars and the increasing fraction of sub-solar-metallicity
stars with increasing projected distance from M104 moving from top to bottom in each panel.     
However, the variation in MDF with projected distance seen in Fig.~\ref{mdfradbinfig} is subtle enough that no statistically significant variations are seen \textit{within} either the UVIS or ACS field in median metallicity or fraction of metal-poor stars.

\subsection{Relationship with Density Profiles \label{densprofsect}}

\begin{figure*}
\gridline{\fig{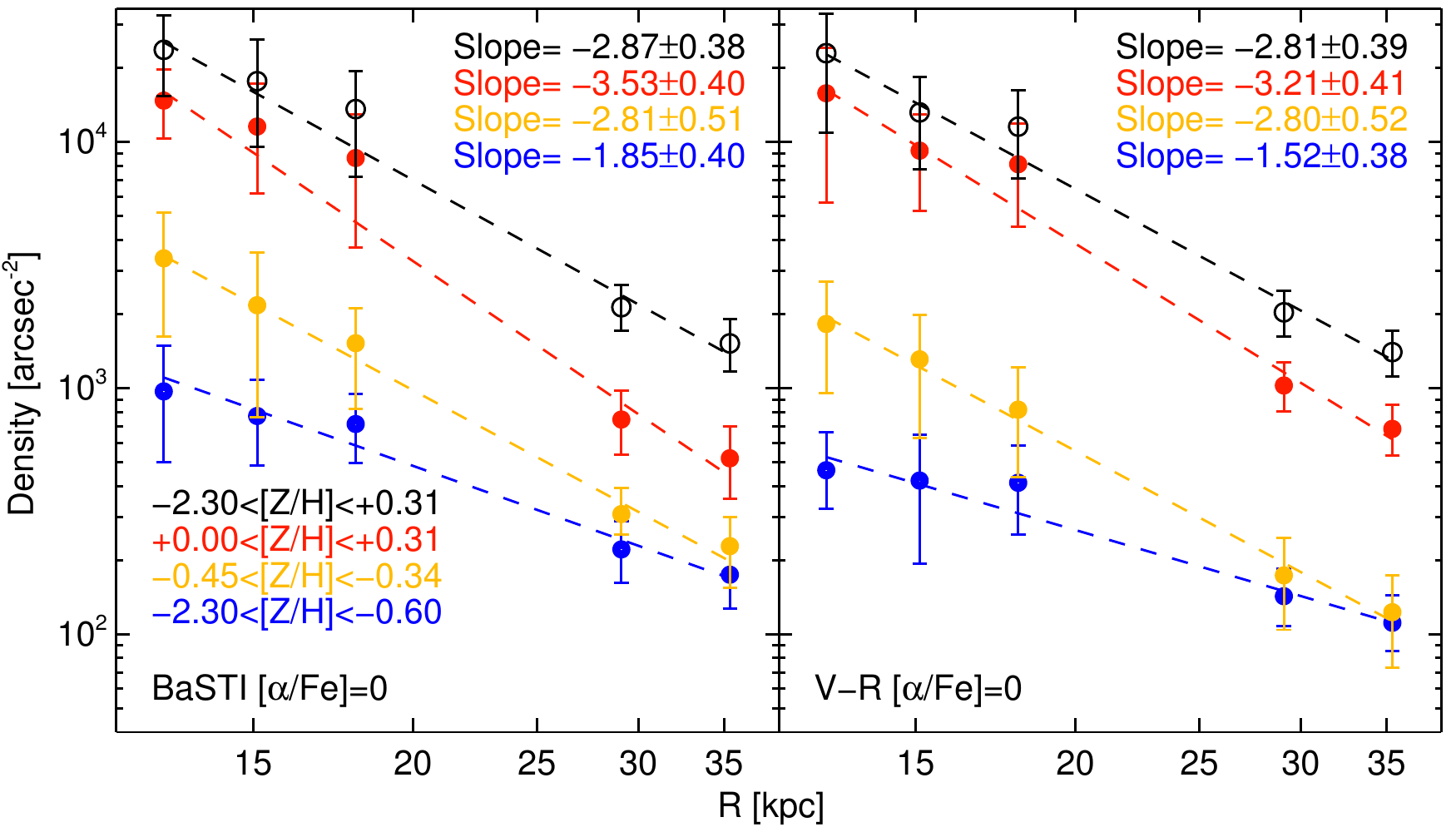}{0.7\textwidth}{}} 
\caption{Radial density profiles for stars over the full sampled metallicity
  range (black open circles) and in three non-neighboring metallicity ranges
  (given in the lower left corner of the left panel and shown as filled circles
  color-coded by metallicity) 
using the same two sets of models as in Figures~\ref{mdfwholefig} and
\ref{mdfradbinfig}, after accounting for incomplete azimuthal coverage.  Power
law slopes best fitting each density profile are overplotted as dashed lines,
and the slopes and their uncertainties are reported in the upper right corner of
each panel.     
\label{densprofilefig}}
\end{figure*}

The results from Figures~\ref{mdfwholefig} and \ref{mdfradbinfig} 
imply the presence of a metallicity-dependent density gradient across
the range of projected radii sampled by the imaging.  To test for
changing projected radial density gradients as a function of
metallicity, we now make the opposite tradeoff as in
Sect.~\ref{sfhwholesect}; namely, we retain the spatial resolution
from the radial bins and search for coarse differences as a function
of metallicity.  To this end, we select three non-neighboring
metallicity ranges and plot their stellar densities in each radial bin
in Fig.~\ref{densprofilefig}.  
We find that for each of the three metallicity ranges, the power law
slopes agree to within their uncertainties across the three isochrone
sets, all finding a metallicity dependence on the power law density
slope such that the more metal-poor population has a flatter density profile
while increasingly metal-rich populations have increasingly steep power law
density profiles. 
Meanwhile, the density profile of the entire stellar sample (with no metallicity
cuts), shown using black open circles in Fig.~\ref{densprofilefig}, has a power
law slope of $\delta$(log $\Sigma$)/$\delta$(log $R_{\textrm{gal}}$)$\sim -2.8
\pm 0.4$.% dex/kpc.
(where $\Sigma$ denotes the projected stellar density).  
If we compare this slope to the
surface brightness profile of \citet{hargis_gcs}, who fit only S{\'e}rsic and
$r^{1/4}$ laws, their data give a power-law slope of $-2.10 \pm 0.01$ 
%dex/kpc 
when fit
over their entire radial range.  However, restricting the fit to only the radial
range occupied by our fields (in practice, only the UVIS field since they
truncate their data at 7 arcmin from the galaxy center), 
their data result in a power-law slope of $-2.70 \pm 0.17$, 
in good agreement with our results.

\section{Discussion \label{discusssect}}

\subsection{Comparison to Other Studies \label{comparesect}}

\subsubsection{Previous Results using WFPC2}

The only other study of the stellar MDF of the Sombrero was presented by
\citet{sombrero_wfpc2}, who studied a field at $\sim$8$\arcmin$ northwest of the
galaxy center using WFPC2 imaging (see Fig.~\ref{fovfig}). They find an MDF
peaking at [Z/H] = $-$0.5 with a tail extending to lower metallicities, and the
authors acknowledge that the most metal-rich population could have been
\textquotedblleft erased by incompleteness\textquotedblright{}.  While they
perform a subset of artificial star tests to measure photometric error and bias
in selected color and magnitude ranges, they find that the bias (mean offset
between output and input color or magnitude) is small and not statistically
significant, and obtain their MDF by 
direct interpolation between 12 Gyr isochrones from \citet{vdb2000}.  However,
based on our discussion of the impact of observational uncertainties on the MDF
in Sect.~\ref{strategysect} illustrated by Fig.~\ref{kernelfig}, it is likely
that the photometry of \citet{sombrero_wfpc2}, which shows significant effects
of incompleteness between 1-1.5 mag faintward of the TRGB in their Fig.~3, may
in fact be susceptible to non-negligible photometric error, bias (and hence
metallicity error distribution) and incompleteness varying as a function of
color \textit{and} magnitude across their CMD.

\subsubsection{Comparison to Other Nearby Galaxies}

\begin{figure*}
\gridline{\fig{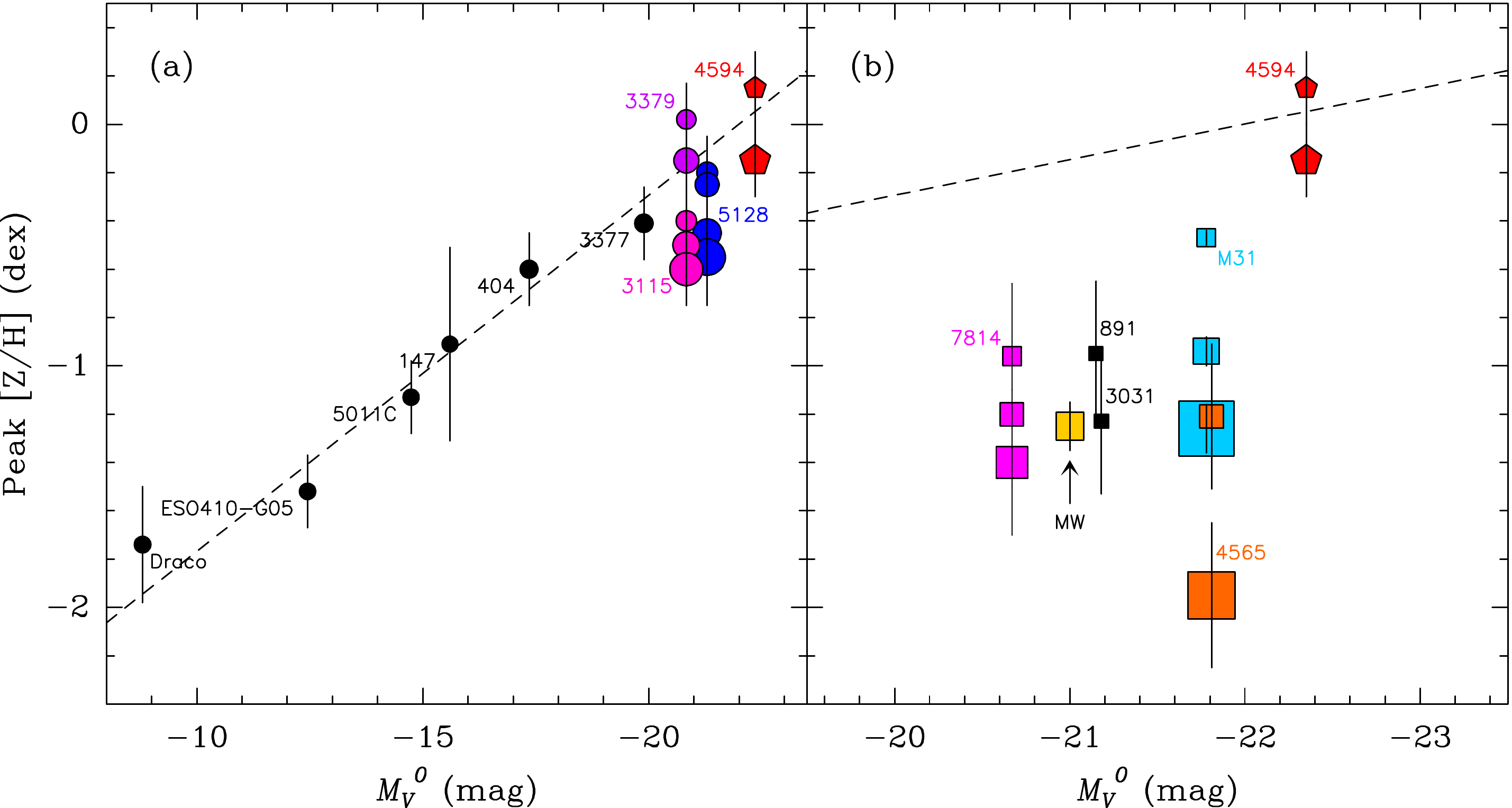}{0.99\textwidth}{}} 
\caption{Peak [Z/H] of the MDF of RGB stars in halo fields around nearby
  galaxies as function of galaxy $M_V^0$ (see Table~\ref{tab:gal_lit}). \emph{Panel (a):} data for early-type galaxies (ETGs). Labels next to data points indicate 
  the NGC number of the galaxy in question (unless indicated otherwise). 
  Small black circles indicate ETG fields with $R_{\rm gal}/R_{\rm eff} \leq 5$, and the dashed line represents a linear least-squares fit to those data.  Symbols and labels with colors other than black  
  represent galaxies with more than one field, and the sizes of those symbols
  scale with $R_{\rm gal}/R_{\rm eff}$ of the field in question. Our two halo fields in the Sombrero are represented by red pentagons.  
  \emph{Panel (b)}: Similar to panel (a), but now showing data for luminous spiral galaxies. Note the different range in $M_V^0$. Small black squares indicate halo fields in spirals for which the dependence of peak [Z/H] on $R_{\rm gal}$ is negligible according to \citet{monachesi+16}. The Milky Way (MW) halo is represented by a yellow square.  The dashed line is the same one as in panel (a).   
\label{fig:ZH_vs_Mv}} 
\end{figure*}

In terms of MDFs of halos of other nearby galaxies using
resolved RGB stars, there are significant differences between spiral
galaxies and early-type galaxies (types E and S0, hereafter
ETGs). Starting with the latter, stellar MDFs of halo fields have been derived
using \emph{HST} data for a handful of giant ETGs, some of which having measurements
for multiple fields at different galactocentric radii. These consist of the S0
galaxy NGC 3115 \citep{peacock_lent_mdf} and the ellipticals NGC 3377
\citep{harris_3377}, NGC 3379 
\citep{harris_3379,jang_3379},
and NGC 5128 
\citep{hh_5128,5128_vimos,rejkuba05,rejkuba_5128,bird_5128_halo}. Some of these
studies also analyzed similar data for nearby lower-luminosity ETGs for
comparison purposes \citep{harris_3379,jang_3379}.  
For halos of spiral galaxies, we use the results of the GHOSTS survey for the
edge-on, ``normal'',  $\sim L^*$ spiral galaxies NGC 891, NGC 3031 (= M83), NGC
4565, and NGC 7814 \citep{ghosts,monachesi+16}, in conjunction
with results for M31 \citep{kalirai+06} and the Milky Way
\citep{ryan91,apogee_halo4}. Relevant data for all these galaxies with
stellar MDF data are listed in Table~\ref{tab:gal_lit}. 
Figure~\ref{fig:ZH_vs_Mv} plots the peak [Z/H] (or mode) of the MDF versus $M_{V}^{0}$ for the galaxies in Table~\ref{tab:gal_lit}; ETGs are shown in panel (a) and spiral galaxies are shown in panel (b). Our results on the Sombrero are shown as red pentagons for comparison. Galaxies with MDF data in more than one field are shown with symbol colors other than black, with the symbol size scaled by the galactocentric radius in units of effective radii of the spheroid (or bulge for the spirals), i.e., $R_{\rm gal}/R_{\rm eff}$.
For reference, a linear least-squares fit to the data of ETGs with MDF data taken in fields with $R_{\rm gal}/R_{\rm eff} \la 5$ is shown as a dashed line, which has a slope of $-0.135 \pm 0.035$ dex/mag,\footnote{Detailed comparisons between
  MDF studies are hampered somewhat by differences in the analyses
  (e.g.,~slightly different CMD regions used to construct the MDF, different
  isochrone models); the latter typically result in systematic differences in
  [Z/H] of $\sim$\,0.10\,--\,0.15 dex. We have adopted 0.15 dex as the minimum
  uncertainty of [Z/H] in Figure~\ref{fig:ZH_vs_Mv}.} 
corresponding to $\mbox{[Z/H]} \propto L_{V}^{0.34 \pm 0.09}$.

\begin{deluxetable}{lrrlrl}
  \tabletypesize{\scriptsize}
\tablecaption{Peak [Z/H] in MDF of RGB stars in halo fields around nearby
  galaxies \label{tab:gal_lit}}
\tablehead{
\colhead{Galaxy} & \colhead{$M_V^0$} & \colhead{$R_{\rm gal}/R_{\rm eff}$} &
 \colhead{$R_{\rm eff}$ ref.} & \colhead{Peak [Z/H]} & \colhead{[Z/H] ref.} 
}
\colnumbers
\startdata
  \multicolumn{6}{c}{\it Early-type Galaxies} \\ [0.2ex]
 UGC\,10822  &  $-8.80$ &  2.\phn  & M12 & $-1.74 \pm 0.24$ & H+07b  \\
 ESO\,410\,$-$\,G005 & $-12.45$ &  1.7 & P+02 & $-1.52 \pm 0.15$ & LJ16  \\
  NGC\,5011C & $-14.74$ &  1.5  & SJ07 & $-1.13 \pm 0.15$ & LJ16  \\
  NGC\,147   & $-15.60$ &  1.\phn   & C+14 & $-0.91 \pm 0.40$ & H+07b  \\
  NGC\,404   & $-17.35$ &  3.5  & B+98 & $-0.60 \pm 0.15$ & LJ16  \\
  NGC\,3377  & $-19.89$ &  4.\phn  & F+89 & $-0.41 \pm 0.15$ & LJ16  \\
  NGC\,3115  & $-20.83$ &  6.7  & F+89 & $-0.40 \pm 0.15$ & P+15  \\
             &  &  14.  & F+89 & $-0.50 \pm 0.15$ & P+15  \\
             &  &  21.  & F+89 & $-0.60 \pm 0.15$ & P+15  \\
  NGC\,3379  & $-20.83$ &  5.4  & F+89 & $+0.02 \pm 0.15$ & LJ16  \\
             &  &  11.8 & F+89 & $-0.15 \pm 0.15$ & LJ16  \\
  NGC\,5128  & $-21.29$ &  7.\phn   & D+79 & $-0.20 \pm 0.15$ & R+14  \\
             &  &  10.5 & D+79 & $-0.25 \pm 0.15$ & R+14  \\
             &  &  15.5 & D+79 & $-0.45 \pm 0.15$ & R+14  \\
             &  &  25.  & D+79 & $-0.55 \pm 0.20$ & R+14  \\
  NGC\,4594  & $-22.35$ &  8.2  & GSJ12 & $+0.15 \pm 0.15$ & C+20  \\
             &  &  17.1 & GSJ12 & $-0.15 \pm 0.15$ & C+20  \\ [0.2ex]
  \multicolumn{6}{c}{\it Spiral Galaxies} \\ [0.2ex]
  NGC\,7814  & $-20.67$ & 11.5 & F+11 & $-0.96 \pm 0.30$ & M+16 \\
             &  & 17.2 & F+11 & $-1.20 \pm 0.30$ & M+16 \\
             &  & 28.7 & F+11 & $-1.40 \pm 0.30$ & M+16 \\
  NGC\,891   & $-21.15$ & 5.\phn  & F+11 & $-0.95 \pm 0.30$ & M+16 \\
  NGC\,3031  & $-21.18$ & 5.\phn  & B+98 & $-1.23 \pm 0.30$ & M+16 \\
  NGC\,4565  & $-21.81$ & 18.\phn  & W+02 & $-1.21 \pm 0.30$ & M+16 \\
             &  & 50.\phn  & W+02 & $-1.95 \pm 0.30$ & M+16 \\
   M\,31     & $-21.78$ & 10.7 & W+03 & $-0.47 \pm 0.03$ & K+06 \\
             &  & 21.4 & W+03 & $-0.94 \pm 0.06$ & K+06 \\
             &  & 60.7 & W+03 & $-1.26 \pm 0.10$ & K+06 \\ 
\enddata
\tablecomments{Galaxies are listed in order of their
  $V$-band luminosities. Column (1): galaxy name. (2): absolute $V$-band
  magnitude from NED. (3) ratio of galactocentric radius of halo field in terms
  of the effective radius of the galaxy, $R_{\rm gal}/R_{\rm eff}$. (4)
  reference for $R_{\rm eff}$: M12 = \citet{mcconn12}, P+02 = \citet{parodi+02},
  SJ07 = \citet{ngc5011c}, C+14 = \citet{ngc147}, B+98 = \citet{baggett+98},
  F+89 = \citet{faber+89}, D+79 = \citet{dufour+79}, GSJ12 =
  \citet{gadotti}, F+11 = \citet{fraternali+11}, W+02 = \citet{wu+02}, and W+03 = \citet{widrow+03}. (5) peak [Z/H] of RGB stars. (6) reference for
  [Z/H] data: H+07b = \citet{harris_3379}, LJ16 = \citet{jang_3379}, P+15 =
  \citet{peacock_lent_mdf}, R+14 = \citet{rejkuba_5128}, C+20 = this paper, M+16
  = \citet{monachesi+16}, and K+06 = \citet{kalirai+06}. }
\end{deluxetable}

A number of relevant conclusions can be drawn from Figure~\ref{fig:ZH_vs_Mv}: 
\begin{itemize}
\item
  Among ETGs, there is a clear correlation between the peak [Z/H] of the halo
  MDF and the $V$-band luminosity (and, most likely, the mass) of the parent
  galaxy when considering fields with $R_{\rm gal}/R_{\rm eff} \la 5$.
  This is
  thought to reflect a manifestation of the galaxy mass-metallicity relation
  \citep[see also][]{gallazzi+05,harris_3379,kirby+13,jang_3379}. 
\item
  Among the sample of nearby galaxies shown in Figure~\ref{fig:ZH_vs_Mv},  
  halos of spiral galaxies are typically much more metal-poor (in terms of the
  peak [Z/H]) than those of ETGs at a given galaxy luminosity. 
  An exception to this is the innermost field of M31 shown in
  Figure~\ref{fig:ZH_vs_Mv}, which \citet{kalirai+06} interpret as being due to a significant contribution of bulge stars.
\item
  Metallicities among $L^*$ spirals (both in terms of peak [Z/H] and its radial
  gradient) span a wide range (see also \citealt{monachesi+16}). This is in
  strong contrast with the situation among ETGs:\ while ETGs with
  $L \approx L^*$ do show negative radial gradients of [Z/H] (see more
  on that below), the peak [Z/H] never reaches values $< -1.0$, even in the
  distant outskirts of their halos (e.g., the outermost fields in NGC 3115, NGC
  5128, and the Sombrero have $17 \la R_{\rm gal}/R_{\rm eff} \la 25$).  
\end{itemize}
Note that in terms of the peak [Z/H] values, the halo fields at
$R_{\rm gal}/R_{\rm eff}$ = 8 and 17 in the Sombrero are roughly consistent
with the trend with galaxy luminosity among 
ETGs shown in Figure~\ref{fig:ZH_vs_Mv}, while they are significantly more
metal-rich than any luminous spiral galaxy halo with MDF measurements studied to date. Interestingly, this includes NGC 7814, a luminous early-type (Sa)
spiral galaxy with a $B/T$ ratio that is very similar to that
of the Sombrero, whereas the peak metallicity of the halo of NGC 7814 is a full order of magnitude lower than that of the Sombrero. This is consistent with the finding by \citet{paul+03} that the globular cluster population of NGC 7814 has a very low fraction of metal-rich clusters for its $B/T$ ratio relative to the Sombrero.

% PG added Sep 2019
Moreover, this also seems consistent with the correlation between \emph{halo} mass and metallicity among $L^*$ spiral galaxies in the GHOSTS survey
\citep{harmsen}, given the high surface number density of RGB stars in the outer halo of the Sombrero relative to those in the GHOSTS survey. 
To check this quantitatively, we convert our surface number densities to halo masses between galactocentric radii of 10 and 40 kpc (hereafter $M_{10-40}$), which \citet{harmsen} found to be a fraction 0.32 of the total stellar halo mass based on theoretical models. We compute this conversion for RGB stars brighter than our adopted magnitude limit (i.e., $M_{\rm F814W} \leq -2.72$) as a function of metallicity using the BaSTI isochrones, adopting an age of 12 Gyr and a \citet{chabrier03} initial mass function (IMF). Initial stellar masses were converted to present-day masses using the \citet{bc03} models.  Under the assumption of a circular halo (see, e.g., GSJ12), we obtain log\,$(M_{10-40}/M_{\odot}) = 10.6$ for the Sombrero, corresponding to a total stellar halo mass of log\,$(M/M_{\odot}) = 11.1$. Repeating these calculations for ages of 8 Gyr and 15 Gyr as well as a \citet{kroupa} IMF, we estimate the uncertainty of this halo mass to be of order 30\% within the framework of the BaSTI models. 

Using the Illustris cosmological hydrodynamical simulations, \citet{illustris}
found that the halo mass-metallicity relation found by \citet{harmsen} mainly
arises because the bulk of the halo mass is accreted from a single massive
progenitor galaxy. 
From our MDF measurements described in Sect.\ \ref{resultsect}, we derive a median metallicity of the Sombrero of [Z/H]~$\sim -0.07$ in the radial range of 10-40 kpc.     
Using the scaling prescriptions of \citet{illustris}, a metallicity [Z/H]~$\sim -0.07$, corresponding to [Fe/H]~$\sim -0.32$ given their assumed [$\alpha$/Fe] = 0.3, 
would imply an accreted stellar mass of $\log\,(M_{\rm acc}/M_{\sun}) \sim 11.0 \pm 0.4$ (cf.\ Figure 5 of \citealt{illustris}). Since this is equal to the total halo mass estimated above from the surface number densities of RGB stars to well within 1$\sigma$, this suggests that \emph{the accretion history of the Sombrero was dominated by a single massive merger event} that occurred several Gyr ago, as opposed to our target fields being dominated by substructure due to recent, minor accretion events.  In this context we note that recent wide-field ground-based imaging revealed a smooth, round halo around the Sombrero \citep{heron}, which is consistent with such a major merger causing a strong gravitational perturbation which led to relatively rapid relaxation. 

In terms of radial gradients of peak [Z/H] among giant ETGs and the Sombrero, 
the decrease of $\sim$\,0.25 dex in moving from the 16 to 33
kpc Sombrero fields turns out to be very similar to the radial gradients in giant
ETGs when expressed in terms of $R_{\rm eff}$ (of the bulge in the case of
the Sombrero, see Section~\ref{comparesect}.3): The radial
gradients are 0.028 dex/$R_{\rm eff}$ for the 
Sombrero compared to gradients of 0.019 dex/$R_{\rm eff}$ (NGC 5128), 0.014
dex/$R_{\rm eff}$ (NGC 3115),\footnote{Note however that this gradient for NGC 3115
  should be considered a lower limit due to incompleteness at
  $\mbox{[Z/H]} \ga -0.4$ in the inner field, see
  \citet{peacock_lent_mdf}.} and 0.026 dex/$R_{\rm eff}$ (NGC 3379). 

On the other hand, a more complex picture emerges when comparing the radial
surface number density profiles of the metal-rich ([Z/H]\,$> -0.8$) and metal-poor
($\mbox{[Z/H]} \leq -0.8$) subpopulations. In most cases, the data show density
profiles that are steeper for the metal-rich stars than for the metal-poor
stars, which is consistent with the more radially extended nature of the
metal-poor globular cluster subsystems in giant ETGs 
\citep[e.g.,][]{bassino+06,peng+08,paul18}.
However, the differences in slope vary between host galaxies: in NGC 3379,
\citet{jang_3379} find density slopes of $\delta$(log $\Sigma$)/$\delta$(log
$R_{\textrm{gal}}$) = $-3.83 \pm 0.03$ %dex/kpc
and $-2.58 \pm 0.03$ 
%dex/kpc 
for the metal-rich and metal-poor stars, respectively.
We find a similar situation for the Sombrero, but with overall flatter density
profile slopes of $-2.9 \pm 0.4$ and $-1.9 \pm 0.4$ for
similar\footnote{The literature studies cited here typically
  make the division between metal-rich and metal-poor stellar populations at
  [Z/H] = $-0.7$.  The quoted value for our density profile slope corresponds to
  [Z/H] $> -0.6$, and if we instead make this division at [Z/H] = $-0.8$, the
  density slope for the metal-poor population flattens to $-1.2 \pm 0.3$.}
metallicity ranges. 
Meanwhile, the profile slopes are very close to one another for NGC 3115
\citep[$-3.0$ vs.\ $-2.7$;][]{peacock_lent_mdf}, and NGC 5128 does
\emph{not} show any significant differences in density 
profiles as a function of metallicity \citep{bird_5128_halo}. 
This indicates that even though there are general trends among the differences
in density profiles (and hence assembly histories) of metal-rich
vs.\ metal-poor subpopulations in ETG halos, there is also significant
galaxy-to-galaxy scatter.

\subsubsection{The lack of RGB Stars with $\mbox{[Z/H]} < -1$}

In summary, the MDF of the halo of the Sombrero shares various characteristics
of other nearby massive ETGs. 
However, unlike any ETG studied to date, we find essentially no stars with
$\mbox{[Z/H]} <-1$ in our two halo fields. Hence, if there is in fact a 
secondary metal-poor peak at $\mbox{[Z/H]} < -1$ as seen, for example, in NGC
3115, NGC 3379, and NGC 5128, then 
it must be located at even larger galactocentric radii.
If we extrapolate the density profiles in Fig.~\ref{densprofilefig}, the
$\mbox{[Z/H]} < -0.8$ population would start to dominate at a projected radius
of $\sim$\,70\,--\,100 kpc from the Sombrero.  
This would be beyond the radius where the halo is indistinguishable from the
background in profiles of ground-based observations, 
which is $R_{\rm gal} \sim 60$ kpc for both the integrated light \citep{burkhead}
and the globular cluster system \citep{rz04}.

% PG moved this sentence here from the previous paragraph (9/25/19)
For NGC 3379 and NGC 5128 (and, interestingly, M31 as well), the transition to a
substantial metal-poor component occurs beyond $\approx 10-15\;R_{\rm
  eff}$ \citep{harris_3379,rejkuba_5128}. 
For the Sombrero, a comparison in terms of $R_{\rm eff}$ rather
than physical radius is complicated by the question of the fundamental nature of
its outer regions: The fits by GSJ12 using only a (S\'ersic) bulge and
(exponential) disk component find $R_{\rm eff}$ = 3.3 kpc and a bulge
ellipticity of 0.42, placing our minor axis fields at $\sim$\,8 and 17
$R_{\rm eff}$.  While our non-detection of a peak at $\mbox{[Z/H]} < -1$ is 
already unusual in this range of $R_{\rm eff}$, the three-component fit of
GSJ12 that include a power-law halo performs significantly better, but
yields a drastically smaller $R_{\rm eff} \approx$ 0.5 kpc (and similar bulge
ellipticity of 0.46), placing even our innermost field beyond 40 $R_{\rm eff}$ 
of the bulge, which would render the non-detection of a metal-poor component
even more unusual.     

\subsection{Comparison to Globular Cluster Properties and Implications for
  Formation Scenarios \label{implications}}

\begin{figure}
\gridline{\fig{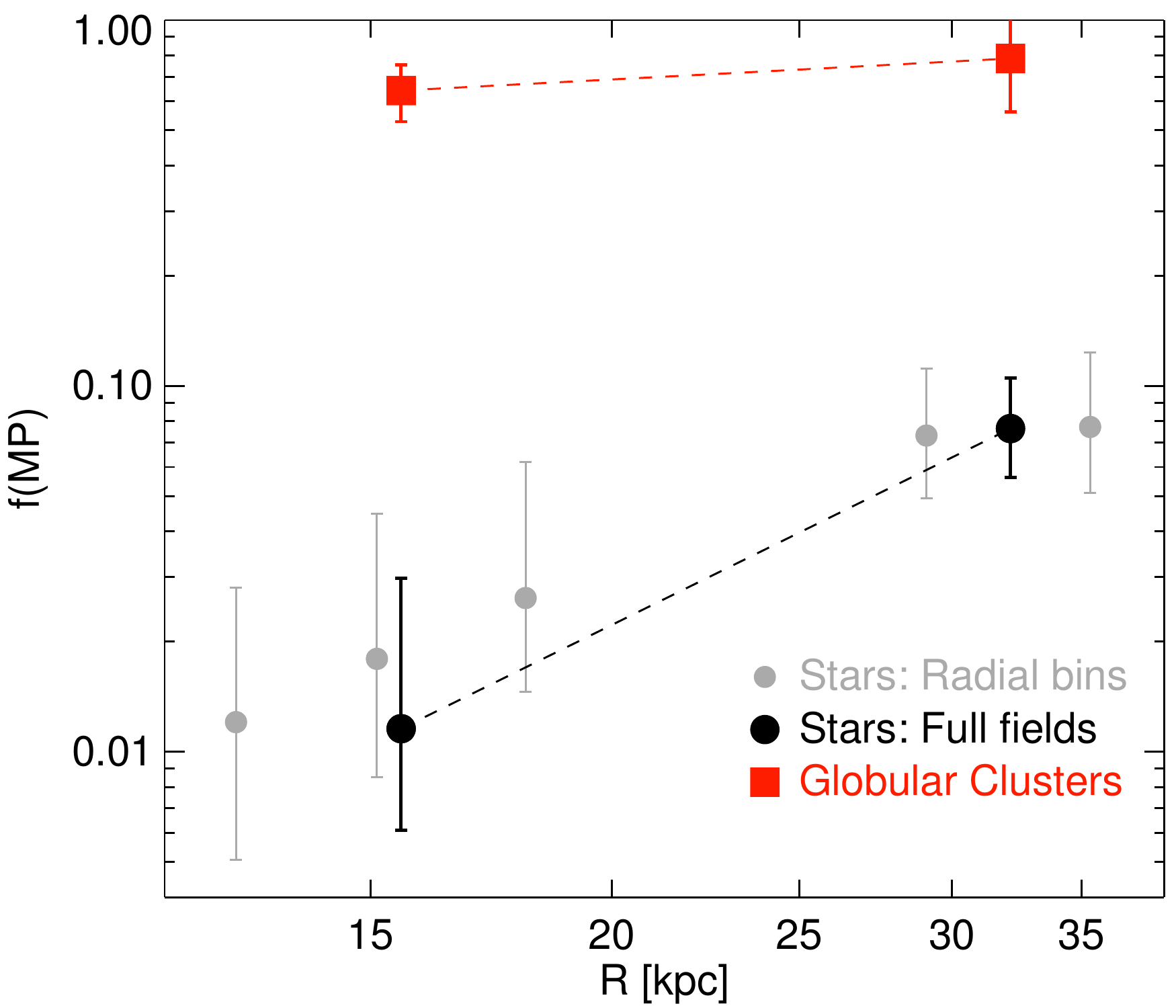}{0.48\textwidth}{}} 
\caption{The fraction of metal-poor ([Z/H] $< -0.8$) stars as a
  function of galactocentric distance, shown in grey for each of the
  individual radial bins (identical to those used in Figs.~\ref{mdfradbinfig}
  and \ref{densprofilefig}) and black for the UVIS and ACS fields in 
  their entirety.  For comparison, the fraction of metal-poor globular
  clusters at the galactocentric distances of the UVIS and ACS fields
  are shown as large red squares.  For clarity, the full-field values
  are connected with dotted lines, and the y-axis is on a logarithmic
  scale. 
\label{frac_mp_starsgcs}} 
\end{figure}

The Sombrero hosts a substantial population of globular clusters (GCs) detected out to
at least 50 kpc, with a significantly bimodal color distribution
\citep{rz04,dowell}. Since the ages of the Sombrero GCs are thought to be old 
(\citealt{larsen02,hempel07}), their colors mainly reflect metallicities,
allowing us to compare radial gradients in the fraction of metal-poor stars
with that of metal-poor GCs.  
Using the extensive GC database provided by \citet{dowell}, we
transform observed $B-R$ colors to [Z/H] according to \citet[][see their
  eq.~4]{hargis_gcs}.  Dividing the sample at [Z/H] = $-0.8$, the fraction of
metal-poor GCs (hereafter $f_{\it GC,\,MP}$) increases from $0.64 \pm 0.11$ to
$0.79 \pm 0.22$ moving radially outward from the UVIS to ACS fields, whereas the
fraction of metal-poor stars ($f_{\it RGB,\,MP}$) increases from
$(1.2\,^{+1.8}_{-0.5})\;10^{-2}$ to $(7.6\,^{+2.8}_{-2.0})\;10^{-2}$ (using the 
BaSTI isochrones) in the same range of $R_{\rm gal}$. 
These trends are illustrated in Fig.~\ref{frac_mp_starsgcs}.  
Such a significant offset between the fraction of metal-poor stars versus
metal-poor GCs in halos of ETGs was also observed in the cases of NGC 3115
\citep{peacock_lent_mdf} and NGC 5128 \citep{hh_5128}.  
However, in the Sombrero, the fraction of metal-poor stars at a given radius is
lower than NGC 3115 by a factor of at least a few (modulo incompleteness at
$\mbox{[Z/H]} \ga -0.4$ in NGC 3115), and a metal-poor ($\mbox{[Z/H]} \la -1$)
peak in the stellar MDF corresponding to the blue GC subpopulation of the Sombrero
\citep{rz04} has yet to be found. 

\begin{figure*}
\gridline{\fig{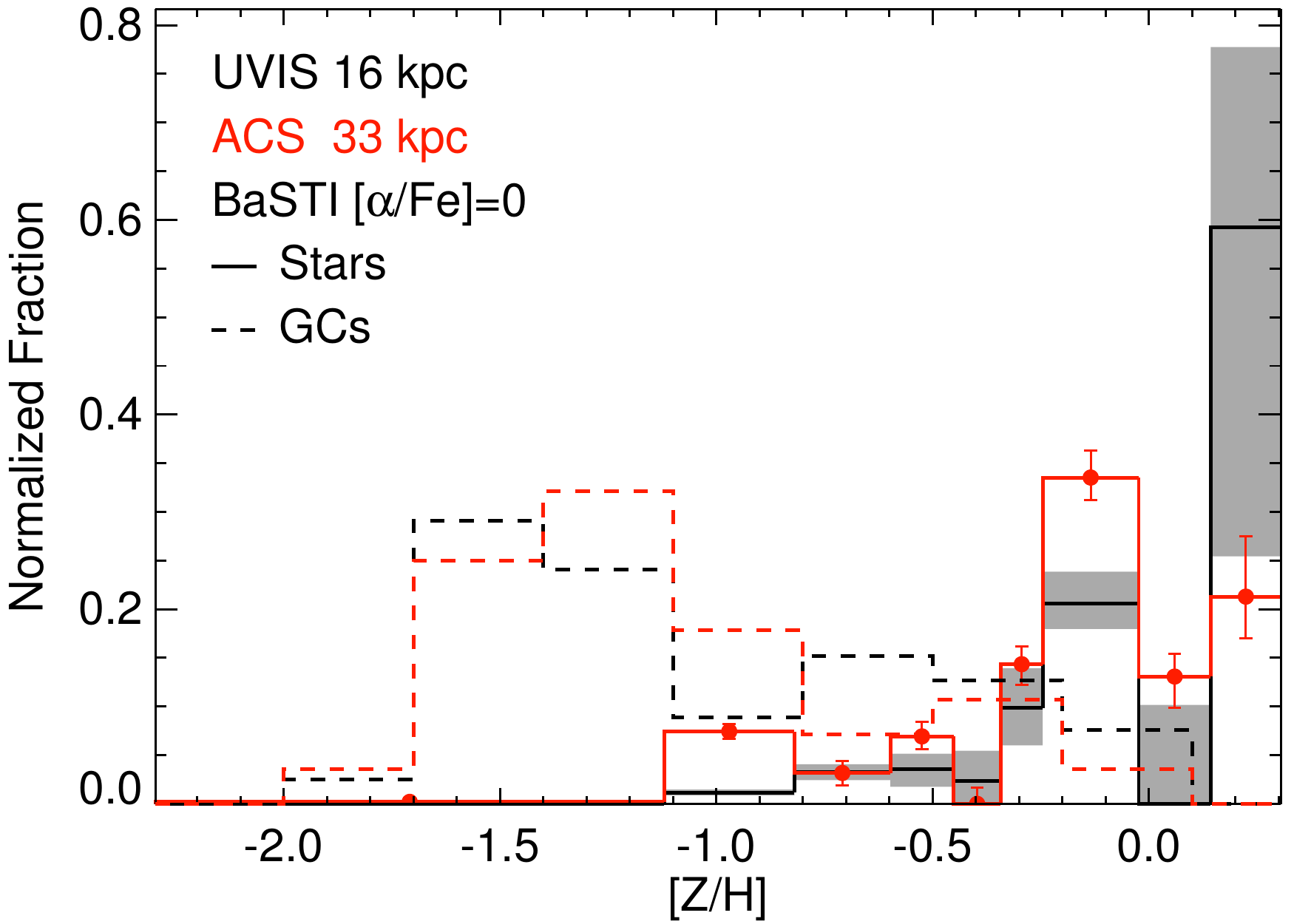}{0.48\textwidth}{}
	  \fig{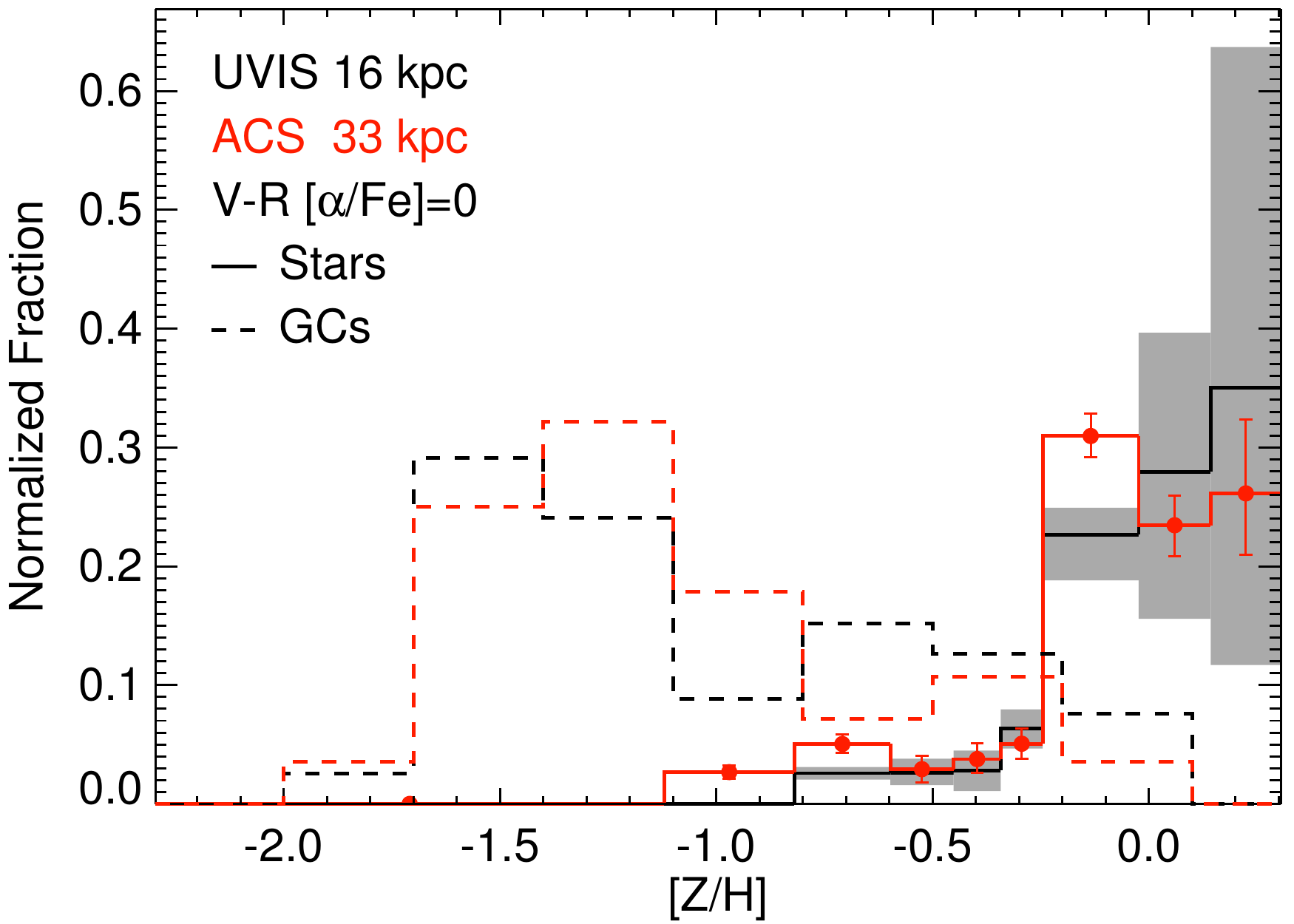}{0.48\textwidth}{}}
\caption{Stellar MDFs as in Figure~\ref{mdfwholefig}, but now with MDFs
  of GCs within identical ranges of projected distance from M\,104 as each
  of our UVIS and ACS fields overplotted using dotted lines (Poissonian error
  bars are omitted for clarity).  The fraction of metal-poor GCs is
  much higher than the fraction of metal-poor stars, discussed in
  Sect.~\ref{implications}.  
\label{mdfgcfig}}
\end{figure*}

These differences have significant implications when considering the dependence
of the specific frequency of GCs ($S_N \propto N_{\rm GC}/L_V$) on metallicity
in the halo of the Sombrero.
In this regard we consider the ratio
  $S_{N,\,{\it MP}}/S_{N,\,{\it MR}}$, where  MP and MR stand for metal-poor 
    ($\mbox{[Z/H]} < -0.8$) and metal-rich ($\mbox{[Z/H]} > -0.8$),
    respectively. 
  This ratio can be expressed as follows:
  \begin{equation}
    \frac{S_{N,\,{\it MP}}}{S_{N,\,{\it MR}}} = \frac{f_{\it GC,\,MP}}{(1-f_{\it
        GC,\,MP})} \; \frac{(1-f_{\it RGB,\,MP})}{f_{\it RGB,\,MP}} \;
    \frac{f_{V,\,{\it RGB}}\,{\rm (MP)}}{f_{V,\,{\it RGB}}\,{\rm (MR)}}
    \label{eq:S_N_ratio}
    \end{equation}
where $f_{V,\,{\it RGB}}\,(Z)$ is the fraction of SSP-equivalent $V$-band
luminosity sampled by the observed RGB stars at metallicity $Z$. 
To determine $f_{V,\,{\it RGB}}\,{\rm (MP)}$ and $f_{V,\,{\it RGB}}\,{\rm
  (MR)}$, we assume a \citet{chabrier03} initial mass function, and use the
BaSTI isochrones to determine $f_{V,\,{\rm RGB}} \,(Z)$, using the bins in [Z/H]
shown in Figure~\ref{mdfwholefig}. 
$f_{V,\,{\it RGB}}\,(Z)$ is known to be strongly dependent on metallicity \citep[see,
  e.g.,][]{goudkrui14}; 
 %Using RGB stars brighter than our adopted magnitude limit (i.e., $M_{\rm F814W} \leq -2.72$), 
we find $f_{V,\,{\rm RGB}} (Z)$ =
0.294, 0.528, and 0.757 for [Z/H] = 0.0, $-$0.8, and $-$1.3, respectively.
From the values of $f_{\it GC,\,MP}$ and $f_{\it RGB,\,MP}$ in the halo fields mentioned
above, we then calculate that
$S_{N,\,{\it MP}}/S_{N,\,{\it MR}}$ $\approx$\,73 and
$\approx$\,23 in the Sombrero's inner and outer halo fields, respectively.  
While the trend of increasing $S_N$ with decreasing metallicity is now
well-established among ETGs \citep[e.g.,][]{hh_5128, peacock_lent_mdf}, the 
ratio $S_{N,\,{\it MP}}/S_{N,\,{\it MR}}$ in the Sombrero is factors of
$\approx 3 - 10$ higher than those seen in other ETG halos.
Moreover, we note that the peak [Z/H] of the metal-poor GC
  subpopulation of the Sombrero is at $\mbox{[Z/H]} \sim -1.5$ \citep{hargis_gcs}, a
  metallicity that is \emph{completely lacking} in the RGB population of the two
  halo fields discussed here. This is illustrated in Figure~\ref{mdfgcfig} which
  compares the MDF of the GCs from \citet{dowell} with that of the RGB stars in
  the two ranges in $R_{\rm gal}$. Since the metal-poor
  GC subsystem populates the full radial extent of the Sombrero's halo, the stellar
  system(s) in which the metal-poor GCs were created must currently have a $S_N$
  value that is significantly higher than any type of galaxy known in the local
  universe.  In this regard, the highest values of $S_N$ among nearby galaxies
  are found in low-mass dwarf ETGs with $-13 \la M_V^0 \la -11$
(\citealt{georgiev+10}; see also \citealt{choksi19}), with $S_N$ values in the
  range 10\,--\,70. Figure~\ref{fig:ZH_vs_Mv} suggests that such galaxies have
metallicities $-1.7 \la \mbox{[Z/H]} \la -1.3$, which is consistent
with the blue GC subpopulation in the Sombrero as well as in giant ETGs
\citep[for the latter, see][]{peng+08,sombrero_gc_phot_3}.

An interesting possibility that can explain (at least in part) 
  the very high $S_N$ value at low metallicity ($\mbox{[Z/H]} \la -1$) for the
  Sombrero, and for giant ETGs in general as well, is that the low-metallicity
  population(s) were accreted at high redshift from progenitors whose properties
  were different from present-day dwarf galaxies in that star formation 
  occurred by means of an initial star cluster mass function
  (ICMF) whose low-mass truncation $M_{\rm min}$ was significantly higher than the
  $10^2\;M_{\odot}$ that is widely adopted in an environmentally independent
  fashion in studies of dynamical evolution of GCs
  \citep[see, e.g.,][]{ladalada03,kruijssen15,li+17}. An increase of $M_{\rm  min}$
  implies a smaller percentage of star clusters that have dissolved completely
  at the present time, thus increasing $S_N$. Recent studies indicate that
  sufficiently high values of $S_N$  result for $M_{\rm min} \ga 10^3\;M_{\odot}$,
  which is expected for high-pressure starbursting environments with turbulent
  velocities $\sigma = 10-100$ km\,s$^{-1}$ and gas surface densities
  $\Sigma_{\rm ISM} \sim 10^2\;M_{\odot}\;{\rm pc}^{-2}$
  \citep{paul18,trujillo+19}, as seen in vigorously star-forming galaxies at
  high redshift \citep[e.g.,][]{tacconi+13}. Future studies of lensed
  high-redshift galaxies with the \emph{James Webb Space Telescope} and future
  30-m-class ground-based telescopes will be able to provide significantly
  improved constraints to $S_N$ and the low-mass regime of the CMF in such
  objects, and hence allow this scenario to be tested.  

\subsection{Future Prospects}

Our results suggest strongly that observations of halo fields in the Sombrero
at even larger galactocentric radii are necessary to fully reveal the nature of
its extended stellar halo.  Both the increasing fraction of
intermediate-metallicity stars at larger radii  as well as the difference in
metal-poor and metal-rich 
density profiles (Fig.~\ref{densprofilefig}) beg for an extended radial
coverage of high-spatial-resolution imaging that is evidently needed to confirm
or refute the presence of a star population with $\mbox{[Z/H]} < -1$. 
Observations at larger radii, less affected by crowding, also open up
two important possibilities for further constraining the formation history of
the Sombrero: First, the presence or absence of substructure in MDF space,
which was detected in 
NGC 5128 by leveraging together an ensemble of HST pointings
probing different galactocentric radii and position angles, yields valuable
clues regarding the nature and importance of past accretion events.  Second,
and with an eye towards forthcoming large-aperture near-infrared imaging
facilities, very deep imaging to measure the red clump properties can place
valuable constraints on the star formation history regarding both the age
distribution \citep{rejkuba05} and the metallicity distribution
\citep[e.g.][]{cohenamiga}. Imaging at longer wavelengths would also improve
our ability to characterize the MDF of the Sombrero at the high metallicities
indicated by this study.

\section{Summary and Conclusions}

We have used deep, high-resolution \textit{HST} imaging of two fields at 16 and
33 kpc projected galactocentric radius along the minor axis of M\,104 to measure
metallicity distribution functions in each field.  Observational effects are
accounted for self-consistently by forward modeling CMDs using artificial star
tests and performing maximum likelihood fits to linear combinations of SSPs at
varying metallicity, assuming a uniformly old (12 Gyr) population.  By analyzing
each field as well as dividing fields into bins in projected radius, we find the 
following results that are robust to assumptions of different evolutionary models:  
 
\begin{enumerate}

\item{The halo of the Sombrero galaxy is strikingly metal-rich, with its MDF
  peaking at near-solar metallicity, consistent with colors from optical
  integrated light.  This is the highest peak metallicity found in halos of
  nearby galaxies to date.  Both of our halo fields show an extreme lack of metal-poor
  stars: $\sim$90\% of the stellar population has [Z/H] $\gtrsim -0.8$, even in
  our outermost field, which has the highest relative fraction of metal-poor
  stars.}   

\item{We detect a gradient in median metallicity of $\delta$[Z/H]/$\delta$(log
  $R_{\textrm{gal}})\sim -$0.5 (or $\sim -0.01$ dex/kpc) from 16 to 33 kpc, and 
  the fraction of metal-poor stars ($\mbox{[Z/H]} < -0.8$) increases by a factor
  of $\sim$\,6 moving radially outward but remains at $\lesssim$\,10\% even in the
  outermost 33 kpc field.}

\item{Based on the recent analysis of the Illustris cosmological
 simulations by \citet{illustris}, we suggest that the metal-rich
 populations in the halo of the Sombrero (with a median $\mbox{[Z/H]} \sim -0.07$) mainly originate from a major merger with a massive galaxy with stellar mass $\log\,(M/M_{\odot}) \sim$~11.1, which is a substantial fraction of the current stellar mass of the Sombrero.} 

\item{The stellar populations of the Sombrero's halo have density profiles
  consistent with power laws at all metallicities, although the data do not
  allow us to discriminate between power law and S{\'e}rsic profiles.  We find a
  global power law density slope of $-2.8 \pm 0.4$, consistent with
  literature surface brightness profiles.  The power law density slope exhibits
  a significant dependence on metallicity, with the super-solar-metallicity
  population falling off with a power law slope $< -3$, while more metal-poor
  ($\mbox{[Z/H]} < -0.8$) stars have much flatter ($> -2$) density profile slopes.}

\item{The fraction of metal-poor globular clusters (GCs) at the projected radii
  of our fields is similar to values seen for other massive early-type galaxies.
  While the existence of a significant offset between the fraction of metal-poor
  globular clusters and metal-poor stars is not unusual, the Sombrero is an
  extreme case in terms of the magnitude of this offset, with the fraction of
  metal-poor stars being lower than that of the metal-poor GCs by approximately
  an order of magnitude, and with stars with $\mbox{[Z/H]} < -1$ being almost
  completely absent, thus resulting in an extremely high specific frequency of
  GCs at those low metallicities. \emph{This lack of stars with $\textit{[Z/H]} \la
    -1$ at the galactocentric distances of our halo fields is unique among
    nearby galaxies} with stellar MDF measurements in their halos. We
  therefore strongly suggest follow-up imaging of additional fields in the outer
  halo of the Sombrero to confirm or refute this apparent lack of a RGB
  population with $\mbox{[Z/H]} < -1$ in the Sombrero. }   
\end{enumerate}

\acknowledgements

We acknowledge useful discussions with Dimitri Gadotti, and we thank the referee for a swift report with insightful suggestions that improved the paper.  Support for program
HST-GO-14175 was provided by NASA through a grant from the Space Telescope
Science Institute, which is operated by the Association of Universities for
Research in Astronomy, Inc., under NASA contract NAS 5-26555.  
OG was supported in part by NSF through grant 1909063.

%\emph{Any other funding acknowledgements?}

\vspace{5mm}
\facilities{HST (ACS/WFC,WFC3/UVIS)}

\appendix

Our observed CMDs and luminosity functions in Figs.~\ref{cmdfig} and
\ref{completelffig} reveal a substantial population of stars lying brightward of
the Sombrero TRGB.  This population is comprised of a combination of photometric
blends and stars scattered brightward of the TRGB by photometric errors
\citep[e.g.][]{harris_3377}, but also, long-period RGB and AGB variables
\citep{renzini_agb,gregg_agb}.  Our observations are clustered in time and are
therefore not ideal for detailed variability studies.  Nevertheless, to gain
some insight on the influence of variables on our observed CMDs and the
resulting MDFs, here we quantify variability fraction as a function of magnitude
in our observed catalogs.  This is done using the individual epoch observations,
normalized to their uncertainties, following \citet[e.g.][]{conroydolphot}.
Specifically, for the artificial stars, which are input as non-variable, the sum
of squared deviations in the time series with respect to the mean magnitude,
denoted as {\it F814W}$_{rms,norm}$ should have an average of 1, so that
$\sqrt{\frac{1}{N_{obs}} \Sigma \frac{\Delta m_{i}}{\sigma_{i}}}$ = 1.

First, we test the hypothesis that the \textit{individual epoch} photometric
errors reported by DOLPHOT are accurate (we know that the total reported
per-filter errors are not, motivating our use of extensive artificial star
tests in our MDF analysis).  If this is the case, then a histogram of {\it
  F814W}$_{\rm rms,norm}$ should have a peak value close to one.  We found that
using the raw, reported individual epoch measurements, {\it F814W}$_{\rm rms,norm}$
for the artificial stars is increasingly skewed to higher values at brighter
magnitudes, suggesting that the individual epoch errors are somewhat
underestimated, and hints of this trend are also apparent in fig.~7 of
\citet{conroydolphot}.  We found that by increasing the individual epoch errors
by $\sim$\,0.025 mag, this trend was mitigated, yielding values of {\it
  F814W}$_{\rm rms,norm}$ centered on 1 for the artificial stars (note that this is
an uncertainty on the \textit{individual epoch} photometric errors, and so has
a negligible effect on the output \textit{mean} magnitudes and errors.
Furthermore, such an increase in individual epoch uncertainties is consistent
with the margin allowed by uncertainties in aperture corrections and the
absolute photometric accuracy of DOLPHOT.  In any case, because candidate
variables in our observed data are identified using a clip on the
\textit{distribution} of output artificial star {\it F814W}$_{\rm rms,norm}$ versus
magnitude, the sample of candidate variables in insensitive to whether we
choose to apply this correction.

\begin{figure}
\gridline{\fig{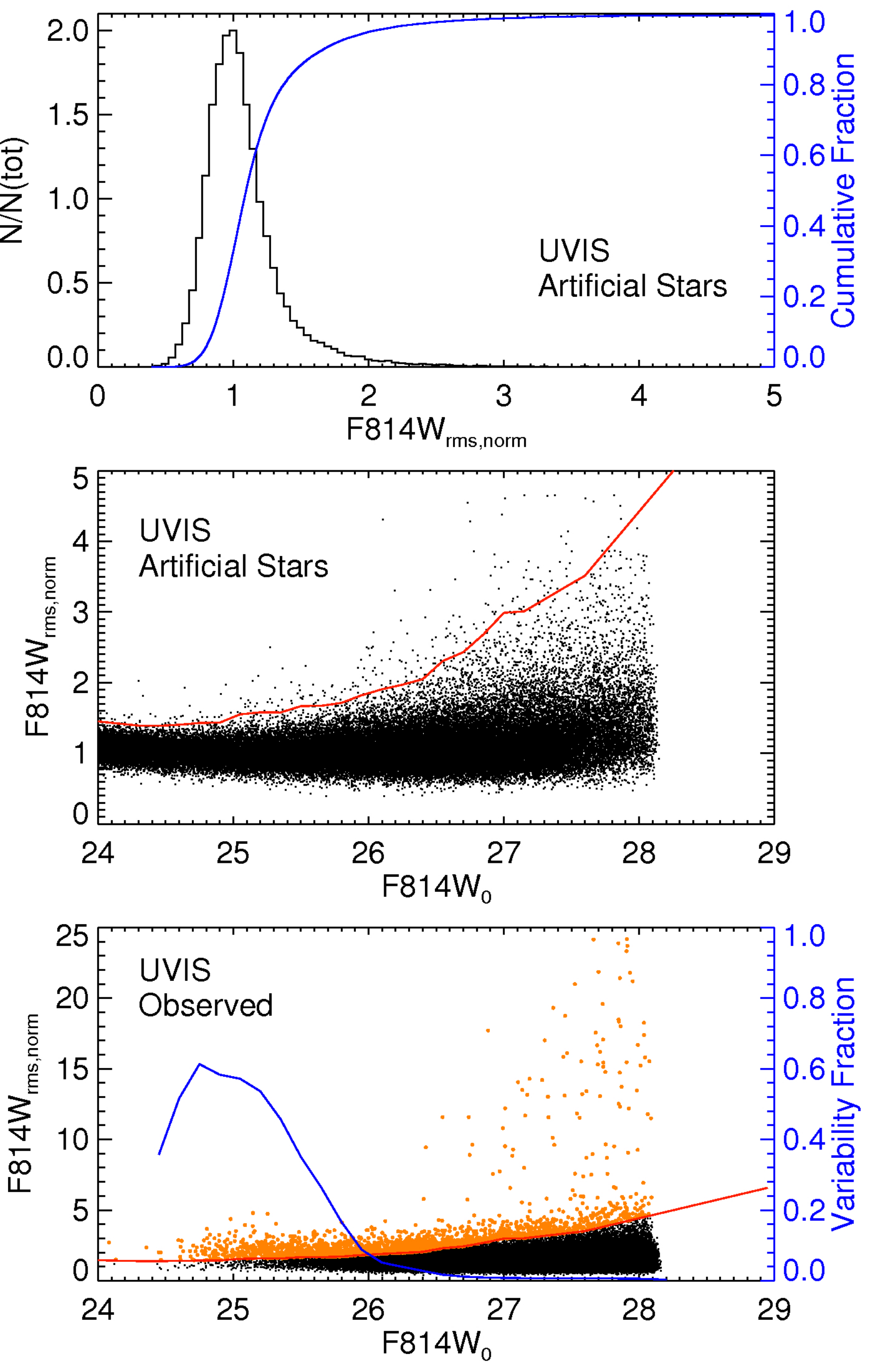}{0.49\textwidth}{}
	  \fig{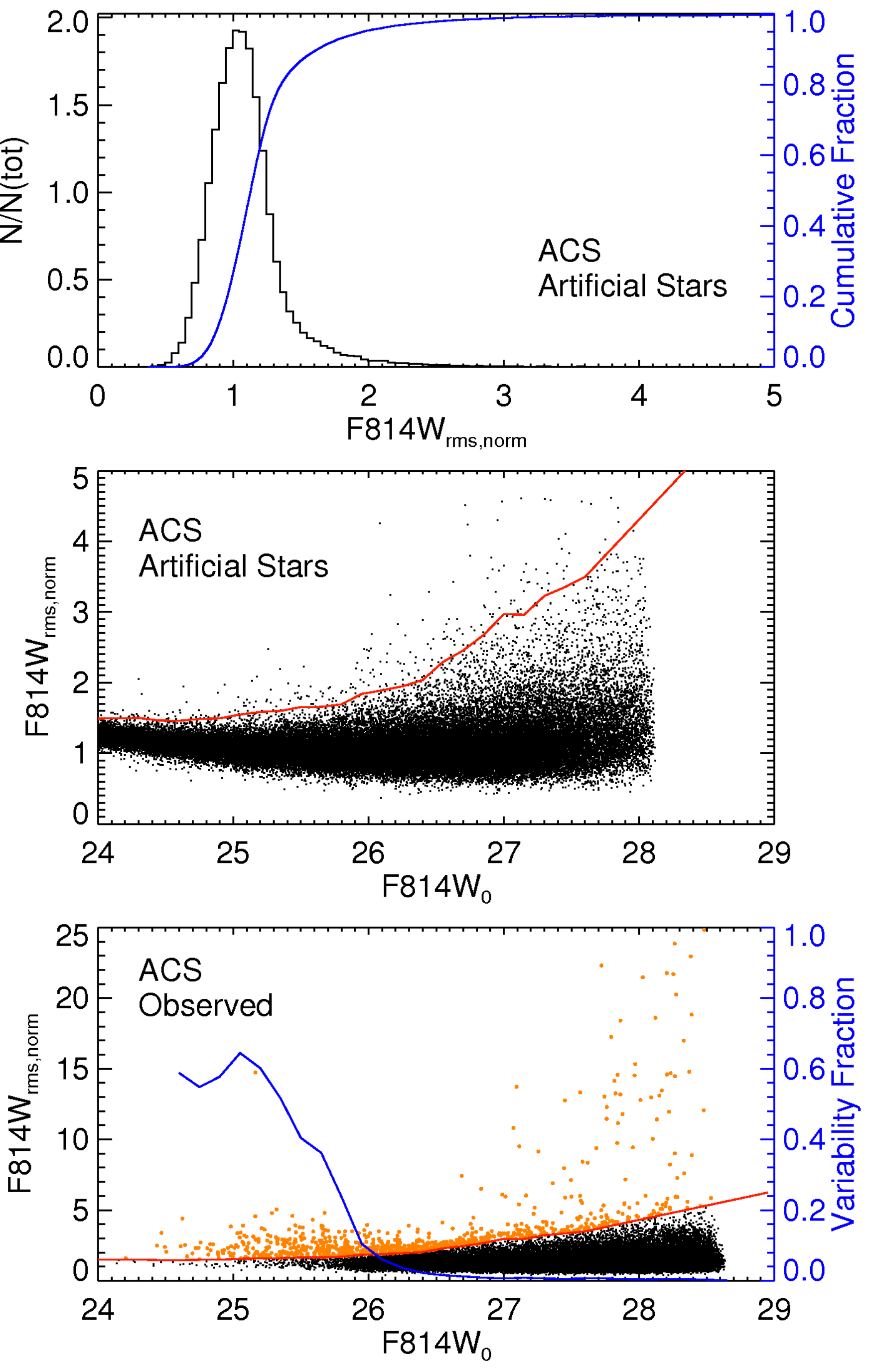}{0.49\textwidth}{}
         }
\caption{For the UVIS (left) and ACS (right) fields: \textbf{Top: }Histogram of
  {\it F814W}$_{\rm rms,norm}$ for input artificial stars which are assumed to
  be constant.  The cumulative distribution of {\it F814W}$_{\rm rms,norm}$ is
  overplotted in blue with respect to the right-hand axis.  Values shown here
  are after adding an additional per-epoch uncertainty of 0.025 mag (see text).
  \textbf{Middle: }Trend of {\it F814W}$_{\rm rms,norm}$ with {\it F814W} magnitude
  for the artificial stars.  Our criterion for defining variables from the
  99.5\% observed envelope is overplotted in red.  \textbf{Bottom:} {\it
    F814W}$_{rms,norm}$ versus mean {\it F814W} for observed stars, with our
  variability criterion overplotted and candidate variables shown in orange
  (note the difference in y-axis scale compared with the middle panel).  The
  variability fraction is overplotted in blue with respect to the right-hand
  axis.  \label{tsartrms}} 
\end{figure}

In Fig.~\ref{tsartrms}, we show for both the UVIS (left) and ACS (right) fields
a histogram and cumulative distribution of ${\it F814W}_{rms,norm}$ from the
artificial stars in the top panel, and in the middle panel, we define a cut
versus magnitude at the 99.5\% interval of the artificial star {\it
  F814W}$_{rms,norm}$ values calculated in sliding magnitude bins of width 0.3
mag, above which observed stars are considered to be candidate variables.  In
the bottom panel, we show the application of this cut to the observed time
series data, and in blue we plot the variability fraction as a function of {\it
  F814W} magnitude.  It is apparent that this variability fraction remains
quite low ($< 10$\%) faintward of the TRGB, but brightward of the TRGB increases
quite rapidly.   

The trend of variability fraction with magnitude supports the idea that at
least half of the stars brightward of the Sombrero TRGB are variables whose
light curves we are sampling stochastically, observing them with too short a
timespan and too limited an observing cadence to measure their mean magnitudes
over their full variability cycles.  This hypothesis is reinforced by
variability studies of upper RGB and AGB stars in Galactic globular clusters,
most notably for the present application including clusters with relatively
high metallicities such as NGC 104 and NGC 5927.  In the few cases where the
necessary time baseline was achieved \citep[e.g.][]{lw05a,lw05b,m5var}, indeed
a large fraction of stars with true mean magnitudes and colors close to the
TRGB are long-period or semi-regular variables with both amplitudes (tenths of
a magnitude or more in $V$) and periods (tens to hundreds of days) consistent
with this idea.

The result based on Fig.~\ref{tsartrms} that $\sim$half of the stars brighter
than the TRGB are variable is a lower limit since variables with amplitudes
failing our {\it F814W}$_{\rm rms,norm}$ criterion may also be scattered above the
TRGB.  To test whether this is the case, we look to other contributions to the
population of stars brighter than the TRGB, which come from two sources:
Contaminants (both foreground Milky Way stars and background galaxies) and
photometric blends scattered brightward.  Foreground and background
contaminants account for $\lesssim$10\% of the sources detected brightward of
the TRGB (see Fig.~\ref{galcontamfig}), and we can use the artificial stars to
estimate the expected contribution due to blends and photometric errors.
Specifically, comparing the RGB stars with magnitudes within 0.5 mag of the
\citet{sombrero_trgb} TRGB (shown as a red line in Fig.~\ref{cmdfig}) and
within the same color limits of $1\leq {\it(F606W-F814W)}_{0} \leq 2.3$ (to reduce
effects of incompleteness) to those with the same color limits lying brightward
of the TRGB ($0.2<{\it F814W}_{TRGB}-{\it F814W}<1.0$), the artificial star tests
predict 175 and 16 stars for the UVIS and ACS fields respectively in the
brighter magnitude range.  This ratio between the two fields is consistent with
the N$_{blends} \propto \Sigma^{2}$ scaling predicted for photometric blends
\citep[i.e.~eq.~4 of][]{harris_3377}.   
Compared to the observed total numbers of stars with
$0.2<{\it F814W}_{TRGB} - {\it F814W} < 1.0$, 661 and 206 in the UVIS and ACS fields
respectively, subtracting the predicted number of blends and
foreground/background contaminants predicts that the majority of sources $> 0.2$
mag brightward of the TRGB are variables.  Furthermore, the gradient in their
projected densities between the UVIS and ACS fields is consistent with a power
law with slope $-1.9 \pm 0.3$, in good agreement with the stellar projected
density gradients shown in Fig.~\ref{densprofilefig}, especially given that our
color cut is biased towards the relatively more metal-poor component.

\end{document}